\documentclass[preprint2]{aastex}
\usepackage{graphicx}

\shorttitle{Tilted torus evolution}
\shortauthors{Janiuk, Proga \& Kurosawa}

\begin{document}

\title{Nonaxisymmetric Effects in the Black Hole Accretion
  Inviscid Hydrodynamics: Formation and Evolution of a Tilted Torus}

\author{Agnieszka Janiuk\altaffilmark{1,2}, Daniel Proga\altaffilmark{1} and 
Ryuichi Kurosawa \altaffilmark{1}}
\altaffiltext{1} {University of Nevada, Las Vegas, 4505 Maryland Pkwy, NV ~89154, USA}
  \altaffiltext{2} {Copernicus Astronomical Center, Bartycka 18, 00-716, Warsaw, Poland}

\begin{abstract}
We report on the fourth phase of our study of slightly rotating accretion 
flows onto black holes. The main new element of this study is that we 
used  fully three dimensional (3-D) numerical simulations.  We consider 
hydrodynamics of inviscid accretion flows. We assume a spherically symmetric 
density distribution at the outer boundary, but brake the flow symmetry by 
introducing  a small, latitude-dependent angular momentum. We also consider 
cases where angular momentum at large radii is latitude- and azimuth-dependent.
For the latitude-dependent angular momentum, 3-D simulations confirm 
axisymmetric results: the material that has too much angular momentum to be 
accreted forms a thick torus near the equator. Consequently, accretion proceeds
only through the polar funnel, and the mass accretion rate through the funnel 
is constrained by the size and shape of the torus, not by the outer conditions. In 3-D simulations, 
we found that the torus precesses, even for axisymmetric conditions
at large radii. For the latitude and azimuth-dependent angular momentum,
the non-rotating gas near the equator can also significantly
affect the evolution of the rotating  gas. In particular, it may
prevent the formation of a proper torus (i.e. its closing, in the azimuthal 
direction). In such models, the mass accretion rate is only slightly less 
than the corresponding Bondi rate. 

\end{abstract}


\section{Introduction}
\label{sec:intro}

Most  galactic nuclei spend a substantial fraction of their lives
in an inactive (``quiescent'') mode. From theoretical point of view,
this inactivity is quite surprising because most galaxies, if not all, contain 
a super massive black hole (SMBH) at their centers
(e.g., Kormendy \& Gebhardt 2001) and a large amount of gas is
available for black hole accretion.
Thus, one would expect vigorous accretion
activity resulting in significant emission of electromagnetic radiation
from all galactic nuclei, but not just from some which are referred to as
active galactic nuclei (AGN).

The modeling of the inactive mode is typically based on the assumption of 
radiatively inefficient accretion
(Ichimaru 1977; Rees et al. 1982; Narayan \& Yi 1994),
in which the rate of accretion  within the radius of influence
of SMBH can be conveniently expressed by
the formula derived by Bondi (1952). 
Some other models focus on the scenario where  
the accretion rate itself is much smaller 
than the Bondi value due to
rotation, magnetic fields, or both, that can lead to convection and 
mass outflows 
(e.g., Begelman \& Meier 1982; Paczy\'{n}ski \& Abramowicz 1982; 
Narayan \& Yi 1995;
Igumenshchev \& Abramowicz 1999; Blandford \& Begelman 1999; 
Stone, Pringle \& Begelman 1999;
Quataert \& Gruzinov 2000; 
Machida, Matsumoto \& Mineshige 2001; 
Hawley \& Balbus 2002; 
Proga \& Begelman 2003b;
Krumholz et al. 2005).

In this paper, we focus on exploring effects of gas rotation on the
black hole accretion hydrodynamics,
using numerical simulations of gas with simplified microphysics .
We assume that  gas accreting onto SMBH is inviscid
and its specific angular momentum ranges from zero to some finite value. 
Such a range of specific angular momentum
is possible for instance in Sgr~A* where there are many
massive stars orbiting the central SMBH
(Genzel et al. 2003; Sch\"odel et al. 2003). As argued by
Loeb (2004), winds from these stars might be a source of 
very low angular momentum gas. Our study is also relevant
to other astrophysical situations, e.g., the collapsar model
for long duration gamma ray bursts (GRBs) where a rotating envelope of
an evolved massive star collapses onto a central compact object
(Woosley 1993; Paczy\'nski 1998; MacFadyen \& Woosley 1999; 
Proga et al. 2003). 

Proga \& Begelman (2003a; hereafter PB03) 
studied the axisymmetric hydrodynamical 
model of the slowly rotating gas.
Here we generalize this model to account for the
non-axisymmetric effects.
This is of a particular interest because the stellar winds at the
Galactic Center are likely to feed the central black hole in a
non-axisymmetric way (e.g., King et al. 2005; Volonteri et al. 2007).
However, non-axisymmetric effects can be important even
for the axisymmetric initial and outer boundary conditions
because of hydrodynamical instabilities.
PB03 showed that a pressure/rotation supported torus forms 
around the black hole and
the accretion rate is smaller than the Bondi rate. 
In the inviscid case, accretion is possible only through
the polar funnels where gas does not rotate or rotates very slowly.
Our goal is to check whether this result holds in 3-D and to what extend
when for example, a non-uniform initial distribution
of the specific angular momentum is assumed.
when there is a gas with zero angular momentum at the equator. 
One can expect that
if the amount of the non-rotating material near the equator is 
large enough, the torus may not form and the accretion rate would not be
much smaller than the Bondi value. 
One of our main questions is whether the accretion can
proceed through the equatorial plane or it can proceed only through the
polar funnels as in an axisymmetric case. 

The content of the article is the following. In Section \ref{sec:method},
we describe the method used in our calculations. 
In Section \ref{sec:results}, we present the simulations results. 
First, we discuss the test runs performed with the
new version of ZEUS-MP code (Sec. \ref{sec:tests}), then, we
describe the results for the initially axisymmetric 3-D case, which is
our 'reference' model (Sec. \ref{sec:reference}), and finally we present
the results for models with non-axisymmetric initial conditions 
(Sec. \ref{sec:main}). 
We discuss our results in Section \ref{sec:diss}.

\section{Method}
\label{sec:method}

In our calculations we use the 3-D code ZEUS-MP (Stone \&
Norman 1992; Hayes \& Norman 2003). 
The code solves the equations of hydrodynamics:

\begin{equation}
{d\rho \over dt} + \rho \nabla {\bf v} = 0
\end{equation}
\begin{equation}
\rho {d {\bf v} \over dt} = -\nabla P +\rho \nabla \Phi
\end{equation}
\begin{equation}
\rho{d \over dt}({e \over \rho}) + P\nabla {\bf v} = 0
\end{equation}
where $\rho$ is the gas density,
 $e$ is the internal energy density, $P=(\gamma-1)e$ is the gas
pressure, and {\bf v} is the velocity of the flow. 
We adopt the ratio of specific heats to be $\gamma = 5/3$. 
We modified the ZEUS-MP code to use the
 pseudo-Newtonian gravitational potential (Paczy\'nski \& Wiita 1980):
\begin{equation}
\Phi(r) = -{G M \over r-R_{\rm S}}
\end{equation} 
where $R_{S}=2 G M/c^2$ is the
Schwarzschild radius.
We start the simulation with the spherical Bondi accretion solution (PB03),
derived iteratively for the density, $\rho(r)$, energy,
$e(r)$, and radial velocity $v_{\rm r}(r)$ 
distributions. 
We express the accretion rate resulting from simulation in the units
of the Bondi accretion rate:
\begin{equation}
\dot M_{\rm B} = \lambda 4 \pi { G^{2} M^{2} \over c^{3}_{\infty}}\rho_{\infty}  
\end{equation}
where $\lambda \approx 0.29$ (see e.g. PB03 for the exact expression
for $\lambda$).

The initial  velocity  $v_{\theta}$
is set zero everywhere, whereas the velocity $v_{\phi}$
is initially non-zero only in a fixed, quasi-conical zone, with a limited
radial size. This zone of initial rotation is formally defined as follows:
\begin{equation}
v_{\phi} = \left\{ \begin{array}{ll}
                 0 & {\rm for}~~\,~~-v_{\rm r}>c_\infty ~~\,~~\\  
     \sqrt{l_{0}} R_{B} c_{\infty} {1-|\cos{\theta}| \over r
  \sin{\theta}}  &  {\rm for} ~~\,~~ -v_{\rm r}<c_{\infty}~~{\rm and}
  ~~~  180^{\circ}-{\Delta \phi_{0}\over 2} < \phi <
   180^{\circ}+{\Delta \phi_{0} \over 2},
\label{eq:phizero}
\end{array}
\right.
\end{equation}
where $l_{0} = 0.1$ is a dimensionless model parameter.
The  other model parameters are: 
$c_{\infty}$,
$\rho_{\infty}$, $M_{\rm BH}$,
$r_{\rm in}$, and $r_{\rm out}$.
The Bondi radius is equal to $R_{\rm B}=
GM/c_{\infty}^{2}$, and in our simulation it is equal to 
1000 $R_{\rm S}$. Our computational
zone extends from 1.5$\times 10^{-3} R_{\rm B}$  to 1.2$ R_{\rm B}$.
When presenting the results, we use the units
of $R_{\rm B}$, $c_{\infty}$, $\rho_{\infty}$ and 
time $t^{'}=t/t_{\rm orb}(R_{\rm B})$. 
We perform our simulations for several values of the parameter
 $\Delta \phi_{0}$.
We use the spherical coordinate system, RTP, and
the boundary conditions in the $r$, $\theta$ and $\phi$ directions are
outflow, reflection and periodic, respectively. 
The resolution in $r$-direction was 140 zones, with $d r_{\rm i+1}/d
r_{\rm i} =1.05$, in $\theta$-direction it was 96 or 100 zones, and 
in $\phi$-direction we had 10, 32 or 60 zones, with
$d \theta_{\rm j+1}/d \theta_{\rm j} = d \phi_{\rm k+1}/d
\phi_{\rm k} = 1.0$. We note here that because of a moderate resolution in the 
$\phi$ direction, the simulations are able to capture only the lowest 
few non-axisymmetric modes.

\section{Results}
\label{sec:results}

\subsection{Test runs}
\label{sec:tests}

Initially, we performed test runs, to check whether
the spherical or axial symmetry is conserved
wherever it should be conserved, and whether the 2-D results of the
axisymmetric calculations are reproduced in 3-D.
Having calculated both the 
models of spherical Bondi accretion ({\bf s}-models), as well as the
axisymmetric accretion with low angular momentum ({\bf l}-models), 
we found that: \\
(a) the results from the 2-D models are reproduced in the $r-\theta$
slices of the 3-D models;
in particular, the accretion rate $\dot M_{\rm in}$ and the flow
pattern are the same;\\
(b) the non-radial velocity components in the {\bf s}-models,
which analytically should be $v_{\theta}=v_{\phi}=0$, can locally 
have non zero values, but both are orders of magnitude smaller than
the local radial velocity and sound speed 
(that ratio is of the order of $10^{-10}-10^{-13}$);\\
(c) the symmetry with respect to both the equatorial plane and 
rotation axis is conserved ($\delta x / x \le 10^{-7}$, where $x$
denotes density, energy, or velocity components) in all models 
at early stages of the evolution, i.e. $t^{'}<0.033$.

The {\bf s}-models conserved their spherical symmetry in both 2-D and
3-D, throughout our runs. For {\bf l}-models, 
as the system evolved, asymmetries began to arise with
respect to the equatorial plane, in both  2-D and 3-D models. 
We note that this was already the case in PB03 simulations.
The axisymmetry in the present 'reference' 3-D model (i.e. the model
with axisymmetric initial conditions) is conserved
initially and at intermediate times of the system evolution 
(however see the results below).

\subsection{Time evolution of the initially axisymmetric flow}
\label{sec:reference}

The computations of an axisymmetric, 2.5-D model of an accretion flow
with low angular momentum were presented in PB03.
We have recalculated this model in 3-D for the purpose of the
present work.
The simulations start from a spherically symmetric 
gas cloud around a black hole, with  density and velocity
distributions derived from the Bondi solution.
The matter located far from the black hole possesses a specific
angular momentum that exceeds the critical value, $l_{\rm crit}=2
R_{\rm S} c$, at the equatorial plane
and is decreasing towards the polar regions (cf. Eq. \ref{eq:phizero}).

The time evolution of the system with the initial conditions 
described above proceeds as follows.
After a transient episode of purely radial infall, when the rotating
material reaches the vicinity of the black hole, a thick torus forms
in the equatorial region. The gas settled in this region is 
supported against gravity by the gas pressure and rotation,
 and the rate of accretion on the black hole, 
$\dot M_{\rm in}$, decreased down to about 30\% of the Bondi 
accretion rate 
(in PB03 the exact value of $\dot M_{\rm in}$ was found to depend
 on the details of the angular momentum distribution). 
This is because the material accretes only through the polar
funnels, while the torus is made of  material that cannot
accrete. (There is no transport due to viscosity; however
the non-axisymmetric shocks can result in some transport of the angular momentum). 
The gas which approached the centrifugal barrier
at the equator, could either outflow radially, or try to turn towards
one of the poles and accrete. Consequently, meridional circulation
movements are observed in the flow.

The accretion rate onto BH varies
in time. Due to meridional circulations, the flow
is not symmetric with respect to the equatorial plane; however,
the time averaged properties (e.g. accretion rate) and the shape of the
torus did settle down to a steady state, as shown in PB03.
In the equator the flow is subsonic down to very small radii, 
while at the poles, at some distance from the center,
the radial velocity exceeds the local speed of sound.

Here, we repeated the calculations of PB03 
in 3-D and ran  new simulations on a relatively long time scale.
Most of our simulation runs lasted up to $t^{'} = 0.36$.
For comparison, the sound crossing time at the $R_{\rm B}$ is about
$t^{'}_{\rm sound}(R_{\rm B}) = 0.125$, while
at the radii corresponding to the dense tori formed in the innermost part of the flow
(see Section \ref{sec:main}),
the sound crossing time is $t^{'}_{\rm sound} (R_{\rm torus})\sim 10^{-3}$.

\begin{figure}
\plottwo{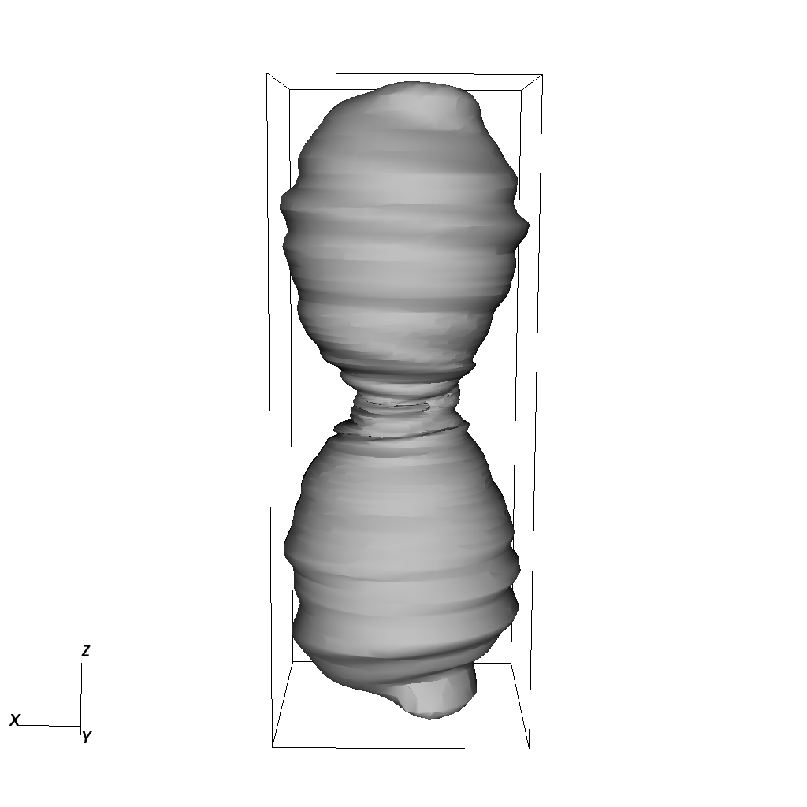}{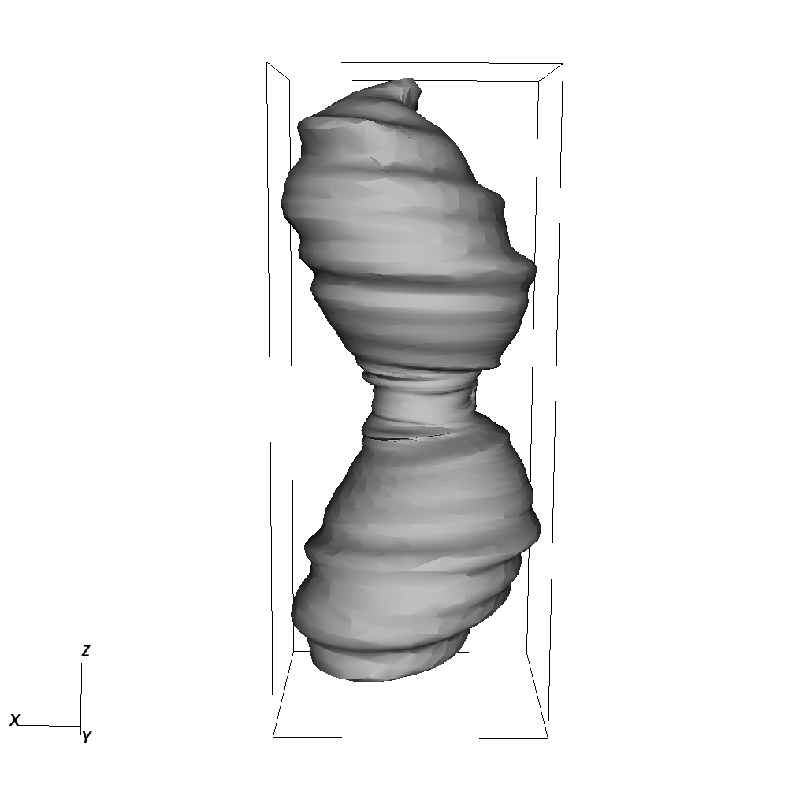}
\caption{ The sonic surfaces for the 3-D models with axisymmetric
  initial conditions (left) and
the non-axisymmetric case 
(right; model $A_{32}$ described in Sec. \ref{sec:main}) at time $t^{'} = 0.1$. 
The extension of the surface in $z$ direction is about 0.065 $R_{\rm B}$, 
while in $x$ and $y$
  direction the box size is about 0.03 $R_{\rm B}$.
}
  \label{fig:sonic3d}
   \end{figure}

Figure \ref{fig:sonic3d} shows the sonic surface, i.e.
the isosurface where the  Mach number, 
$M_{\rm tot}=\sqrt{(v_{r}^2+v_{\theta}^2+v_{\phi}^2)/c_{s}^{2}}$ is a unity. 
at the $t^{'} = 0.1$.
For the zero-vorticity flow,
 the surface shape can be derived analytically, as it passes
 orthogonally through the velocity equipotential surfaces 
(Papaloizou \& Szuszkiewicz 1994). In general, the shape of the surface
can be more complex. As one can see in the Figure, 
the sound waves are propagating outwards in the flow.
The sonic surface for initially axisymmetric model is shown in the left panel,
and for comparison in the right panel we also show the
non-axisymmetric case (model $A$, to be described in
Sec. \ref{sec:main}).

We find that both qualitatively and quantitatively the initial results in
3-D axisymmetric case are
the same as in 2.5-D. The main new feature of the torus in 3-D, 
which could not be studied
in the PB03 simulation, is that in the very late stages of the
evolution ($t^{'} > \sim 0.3$)
the axisymmetry breaks and the inner torus becomes tilted with respect to the
equator, and in the end it starts precessing.
This departure from the symmetry at later times is caused by
the equatorial outflow and meridional circulations in the torus become
suppressed, while the material tries to get through towards 
the black hole and accrete along one of the poles. At the same time, 
for some regions (e.g., $\phi \sim 0^{\circ}$) the accretion occurs
through the northern pole, at the opposite side ($\phi \sim 180^{\circ}$)
and the flow chooses rather the southern pole.
Consequently, a torque is induced and the rotation axis of the torus
changes in time.

\begin{figure}
\plottwo{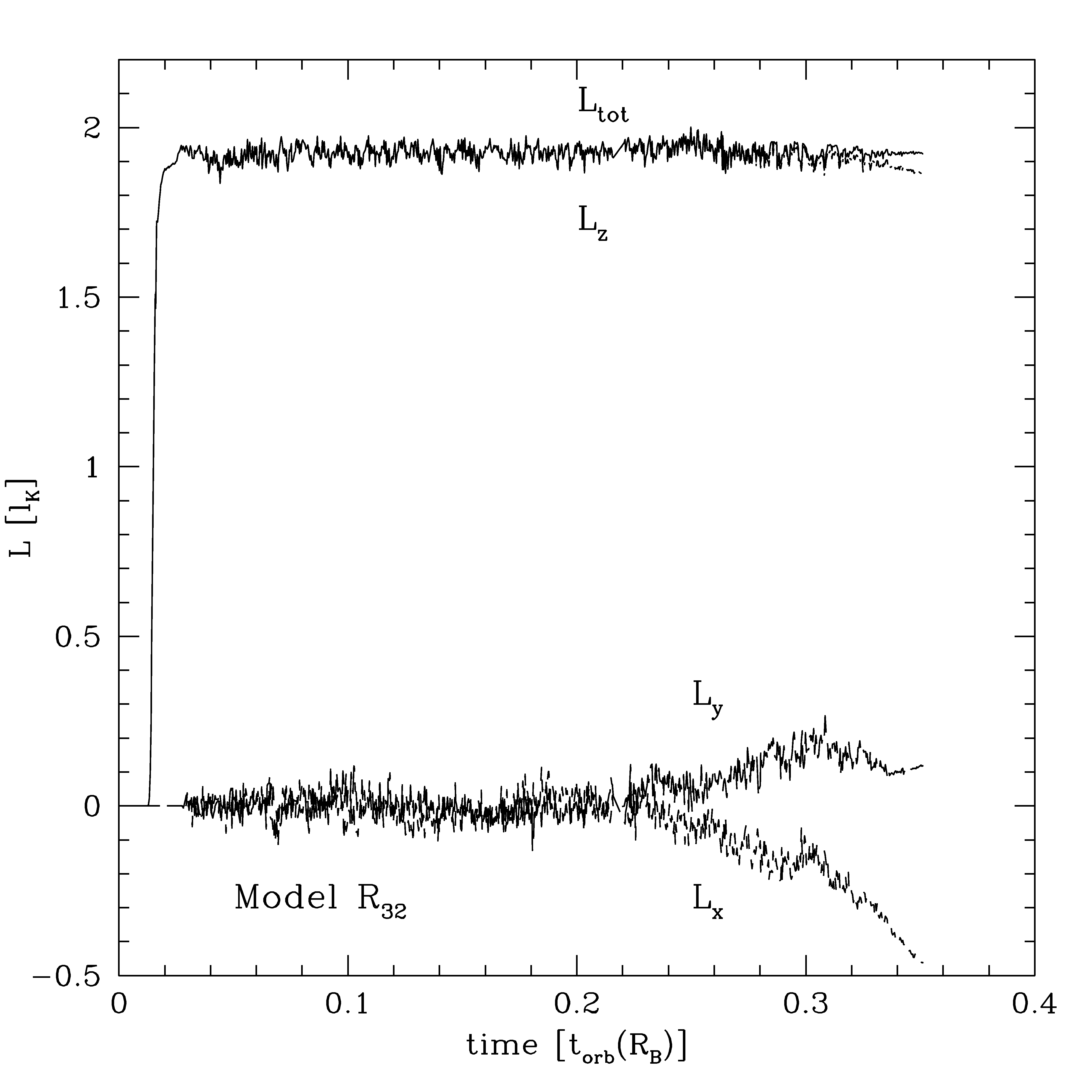}{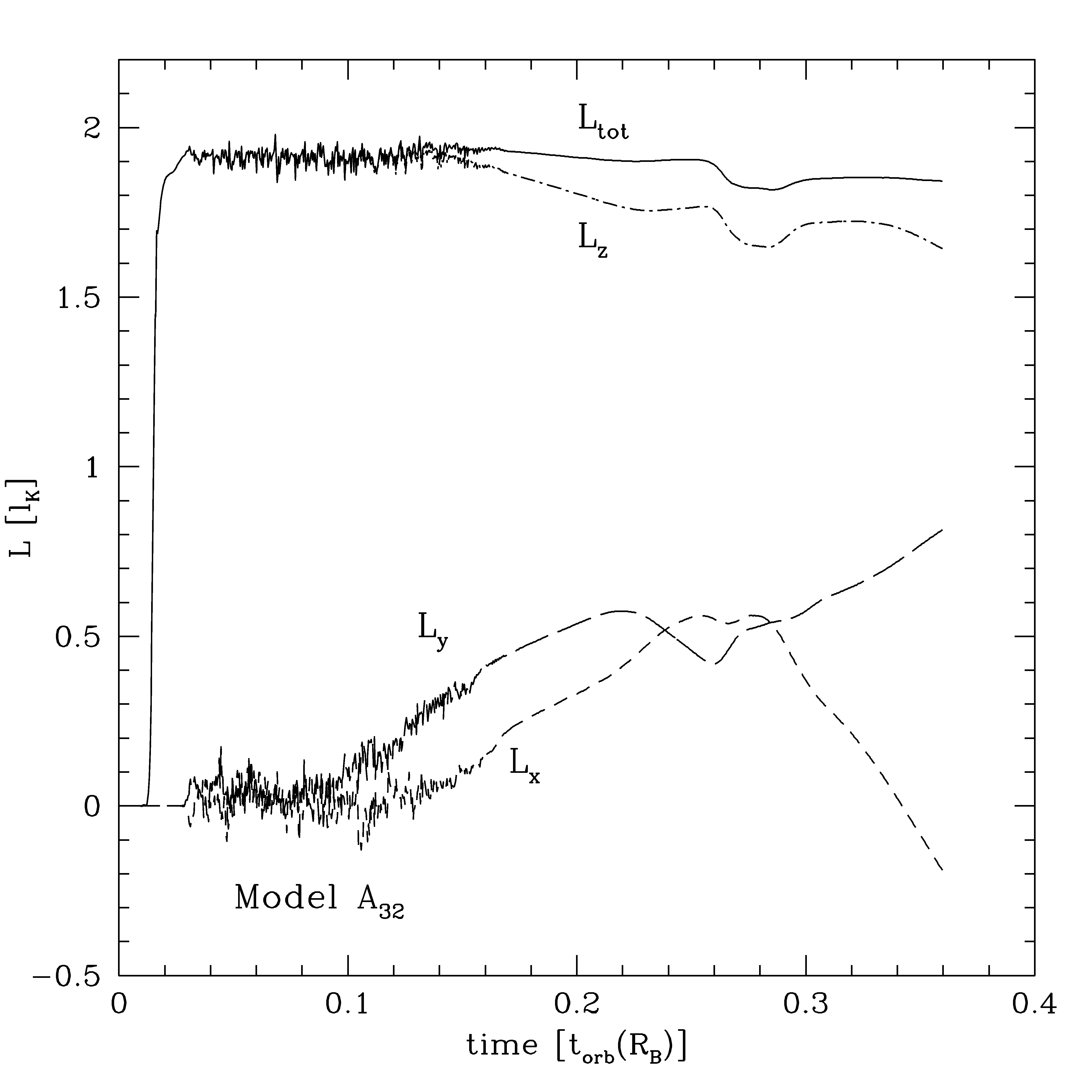}
\caption{The total angular momentum evolution in the initially axisymmetric
  (left) and non-axisymmetric (right; model $A_{32}$ described in
  Sec. \ref{sec:main}) models. The angular
  momentum is in the units of Keplerian angular momentum on the inner
  radius ($l_{\rm k}$, 
and the plot shows the magnitude of {\bf $L$}$_{\rm tot}$
  (solid line) and its components: $L_{\rm x}$ (short dashed line),
$L_{\rm y}$ (long dashed line) and $L_{\rm x}$ (dot dashed line).
}
  \label{fig:levol}
   \end{figure}

The precession of the torus is illustrated in Figure \ref{fig:levol} in
terms of the total angular momentum, {\bf $L$}$_{\rm tot}$, 
which is changing in time.
This figure shows the motion of the innermost part of the flow, i.e. the
torus, defined by the density threshold $\rho_{\rm min}=500$. 
Initially, the dominant component is $L_{\rm z}$, while 
$L_{\rm x}$ and $L_{\rm y}$ are close to zero (but fluctuating). 
It means that the torus rotates basically around the $z$ axis.
After $t^{'} \sim 0.21$, $L_{\rm x}$ and $L_{\rm y}$ are non-zero, 
and the rotation axis tilts towards the 4${\rm th}$ quarter in
the $x-y$ plane. At $t^{'}\sim 0.3$, $L_{\rm y}$ starts 
decreasing while $L_{\rm x}$ still decreases and $L_{\rm z}$ is almost 
constant. It means that 
the torus is now precessing, i.e. the rotation axis 
moves counter-clockwise with  respect
 to an observer along the +z axis.
Figure \ref{fig:levol} also shows for comparison,
the results for the torus
precession in case of  non-axisymmetric initial conditions (model
$A$).
This model will be discussed below in more detail (Sec. \ref{sec:main}).

\begin{figure}
\plotone{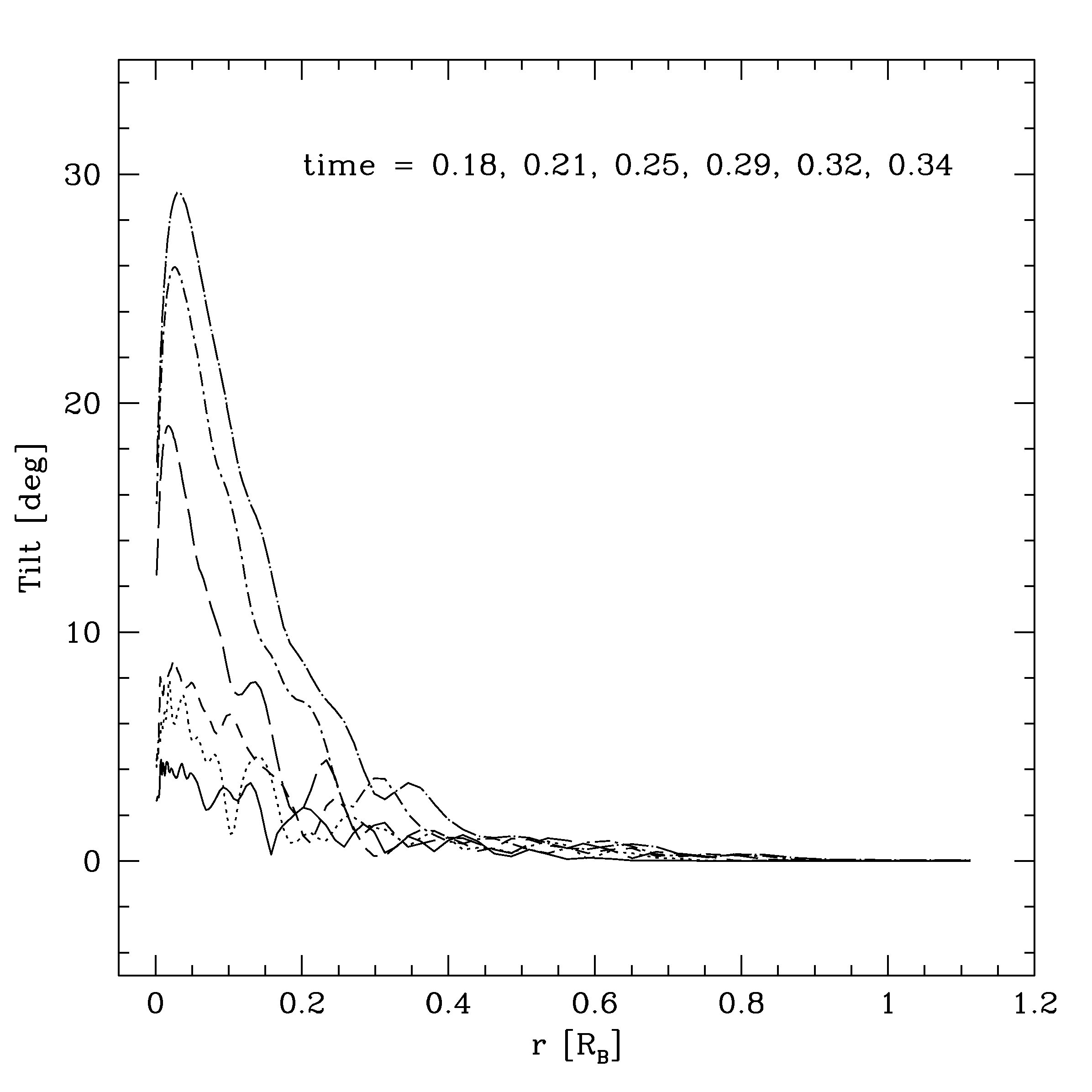}
\caption{The tilt angle $\beta$ (Eq. \ref{eq:tilt}) as a function of radius,
for various times in the late phase of the evolution:
$t^{'}$ = 0.18 (solid line), 0.21 (dotted line), 0.25 (short dashed line), 
0.29 (long dashed line), 0.32 (dotted-short dashed line)  and 0.34 
(dotted-long dashed line).
The initial tilt was zero (initially axisymmetric model). 
}
 \label{fig:tilt}
   \end{figure}

\begin{figure}
\plotone{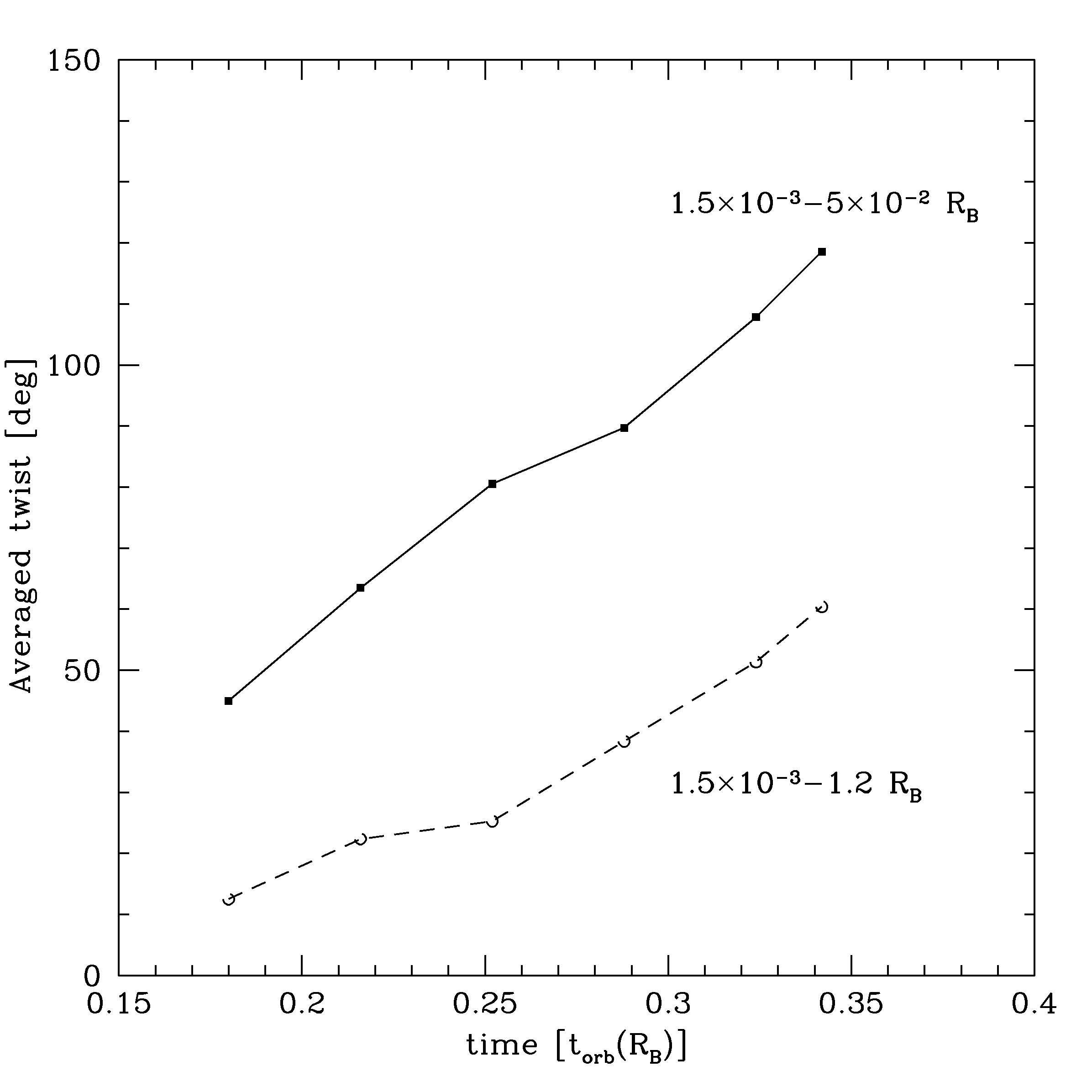}
\caption{The cumulative twist angle $\gamma$ (Eq. \ref{eq:twist}) as a function of time,
for  the late phase of the evolution in the initially axisymmetric model. 
The solid line shows the twist averaged over innermost radii, from
$1.5\times 10^{-3}$ to $5\times 10^{-2}$ $R_{\rm B}$, while the
dashed line shows the twist averaged over the whole disk,
up to 1.2 $R_{\rm B}$.
}
 \label{fig:twist}
   \end{figure}

The precession is also shown in Figures \ref{fig:tilt} and \ref{fig:twist}
(see e.g. Fragile et al. 2007). The tilt, defined as an angle
between the angular momentum vector of the gas and the $z$ axis, is initially equal 
to zero for all radii. Later during the simulation the tilt rises strongly
in the inner parts of the flow. The tilt is equal to:
\begin{equation}
\beta(r,t) = \arccos\Big({L_{z} \over L}\Big)
\label{eq:tilt}
\end{equation}

In the Figure \ref{fig:tilt} we plot the tilt angle, 
$\beta(r,t)$,
as 
a function of radius, for several time snapshots, starting from $t^{'}=0.18$.
The differential form of the torus precession is also visible in
 Figure \ref{fig:twist}, in which we plot the cumulative twist angle, $\gamma(r,t)$,
 as a function 
of time. 
The twist angle is defined as a cumulative angle by which the angular momentum 
vector revolves in the $x-y$ plane, by the time $t$:
\begin{equation}
\gamma(r,t) = \arccos\Big({L_{x} \over \sqrt{L_{x}^{2}+L_{y}^{2}}}\Big)
\label{eq:twist}
\end{equation}
Before the disk was tilted, it did not precess, and by definition 
the twist was zero. Therefore the results in the Figure are also plotted from time 
 $t^{'}=0.18$. The solid line is the twist averaged over the radius for the inner
part of the flow ($1.5\times10^{-3} R_{\rm B} < r < 5\times 10^{-2} R_{\rm B}$), 
i.e. the torus, while the dashed line shows the twist averaged for the
whole range of radii. Clearly, the innermost torus precesses much stronger than the rest of the flow, and the maximum twist at the end of the simulation was $\gamma \approx 120^{\circ}$.

To check whether the tilt and precession of the inner torus found in this simulation 
is a physical or rather numerical effect, we performed several further test simulations.
First, we reversed the direction of the flow in the initial condition, 
i.e. we changed the sign of the azimuthal 
velocity $v_{\phi}$. In this simulation, we also found that the torus tilts and 
starts precessing, at the same time ($t^{'}>0.18$). However, the tilt and precession 
are in the opposite directions as measured with respect to the grid, i.e. the initial 
tilt was $\beta = 180^{\circ}$ and decreased to about $\beta = 150^{\circ}$, while 
the twist was negative.

Second, we checked how transient the effect of precession is.
Due to technical limitations, we were not able to run all the simulations 
for very long time, but we completed one run up to $t^{'}=0.8$, 
for the model $R_{32}$. 
In this model we found that
the tilt angle increased from  $\sim 30^{\circ}$ to  $\sim 40^{\circ}$, and the
tilt spreaded to larger radii, so that not only the innermost parts of the flow
precessed. However, at this very late phase, 
some of the ring-like structures of the largest 
density were broken into two separate parts, each of them of a ``C'' shape. 
Therefore the precession may not be a long-term effect. We plan
to investigate this in near future.

Third, we checked for the importance of the adopted physics of the model, 
in particular, the ratio of the  sound speed to the free fall velocity. 
We calculated two models with much smaller Bondi radius, 
$R_{\rm B}=100 R_{\rm S}$ and $R_{\rm B}=300 R_{\rm S}$, which correspond to 
a much larger sound speed at infinity: respectively, 3.15 and 1.82 times larger than
in all the other simulations.
These models are 
denoted as $M_{100}$ and $M_{300}$ in Table \ref{tab:models}.
The models could be run for much longer time 
in terms of $t_{\rm orb}(R_{\rm B})$, i.e. for $t^{'}=11$ and $t^{'}=2$, 
respectively. However, 
in these models we did not find any signatures of tilt or precession. 
This is because here
the Mach numbers are never large: they are at most $M=3.3 - 3.5$ at the inner radius, 
whereas for the precessing torus the Mach numbers reached the values
as large as 4.7 - 6.0. 
The small Mach numbers in models $M_{100}$ and $M_{300}$ 
make the shocks smaller; hence they do not 
amplify the asymmetries growing out of initial perturbations. 

Finally, we tested the role of the artificial viscosity, which 
might help to 
spread the shocks and avoid precession if it was a numerical artifact.
The artificial viscosity was parametrized with the standard
Neumann-Richtmeyer artificial viscosity coefficient {\it qcon} =2.0 However, 
the results in this simulation were very similar to the original simulation. 
Specifically, for time $t^{'}=0.324$ the maximum tilt angle was $26^{\circ}$, while 
in the former case and $28^{\circ}$ in the latter.

From the above tests, we conclude that the precession in our initially axisymmetric 
model is rather a physical than numerical effect, and is connected with relatively 
large supersonic speeds of the flow achieved in our model.

\subsection{Time evolution for the non-axisymmetric initial conditions}
\label{sec:main}

Now we investigate how the non-axisymmetric initial distribution of the
specific angular momentum affects the evolution of the flow.
In particular, we check whether the 
rotationally supported torus  forms and if a steady state 
can be achieved (with or without torus precession).
We start our simulation with the non-zero specific 
angular momentum enclosed
in a conical region of a width $\Delta \phi_{0}$ 
(see Eq. \ref{eq:phizero}).
A naive prediction could be that the rotating
gas will reach the innermost regions, spiral in, and
after a few orbital cycles
the material with large angular momentum 
would be mixed with the non-rotating gas. 
Therefore a rotationally supported torus would form,
regardless of $\Delta \phi_{0}$.
The only dependence
 on this parameter would be the moment when such a torus forms.
However, as we show below, the numerical simulations lead us to a
different result: depending on $\Delta \phi_{0}$ the rotating gas
may not form a torus at all.

We performed the runs for non-axisymmetric initial conditions 
for a range of $\Delta \phi_{0}$.
The models are summarized in Table 
\ref{tab:models}. The non-axisymmetric models are labeled with the
letters $A$-$E$, while the reference model is labeled as $R$.
As we mentioned in Sec. \ref{sec:method} and as 
the Table shows, we tested the models with smaller (32 zones) and
larger (60 zones) resolution in the $\phi$-direction. We checked,
that the time averaged results for the accretion rate only very weakly
 depend on the resolution, however the amplitude of time variability 
of $\dot M$ increases with resolution.

Below, we present these results for the largest adopted value of $\Delta \phi_{0}=330^{\circ}$ (model $A$). 
This parameter translates into a small non-axisymmetric perturbation
in the initial conditions, i.e.
small content of non-rotating material.

\begin{figure}
\includegraphics[scale=0.7]{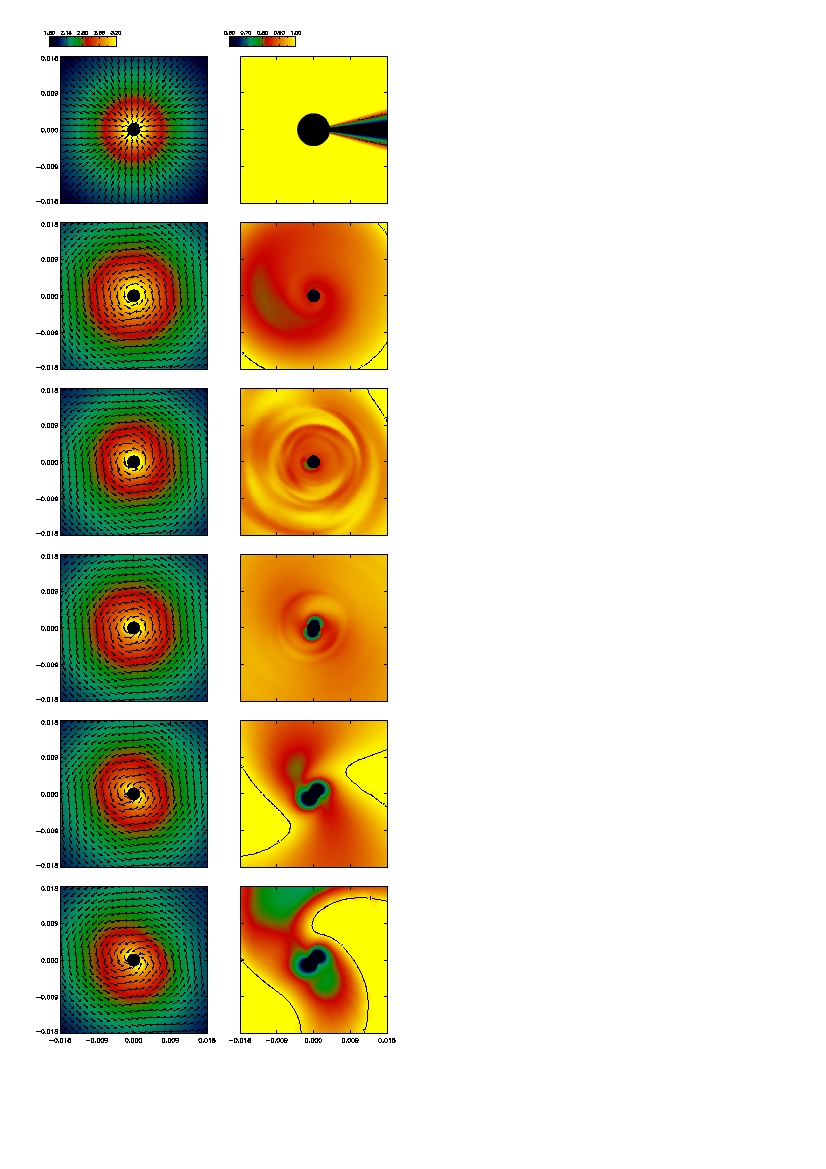}
\vspace*{-28mm}
\caption{The results for model $A_{32}$(see Tab. \ref{tab:models}), for
6 time snapshots, from top to bottom: $t^{'}$ = 0.0, 0.018, 0.09, 0.16, 0.23 and 0.29.
The maps show the central region (only upper right map is zoomed-out) 
in the equatorial
  plane ($\theta=90^{\circ}$): density and velocity field (left) 
and specific angular
  momentum (right). The solid line marks the contour of $l_{\rm
    spec}=l_{\rm crit}$.
}
  \label{fig:torus1}
   \end{figure}

\begin{figure}
\includegraphics[scale=0.7]{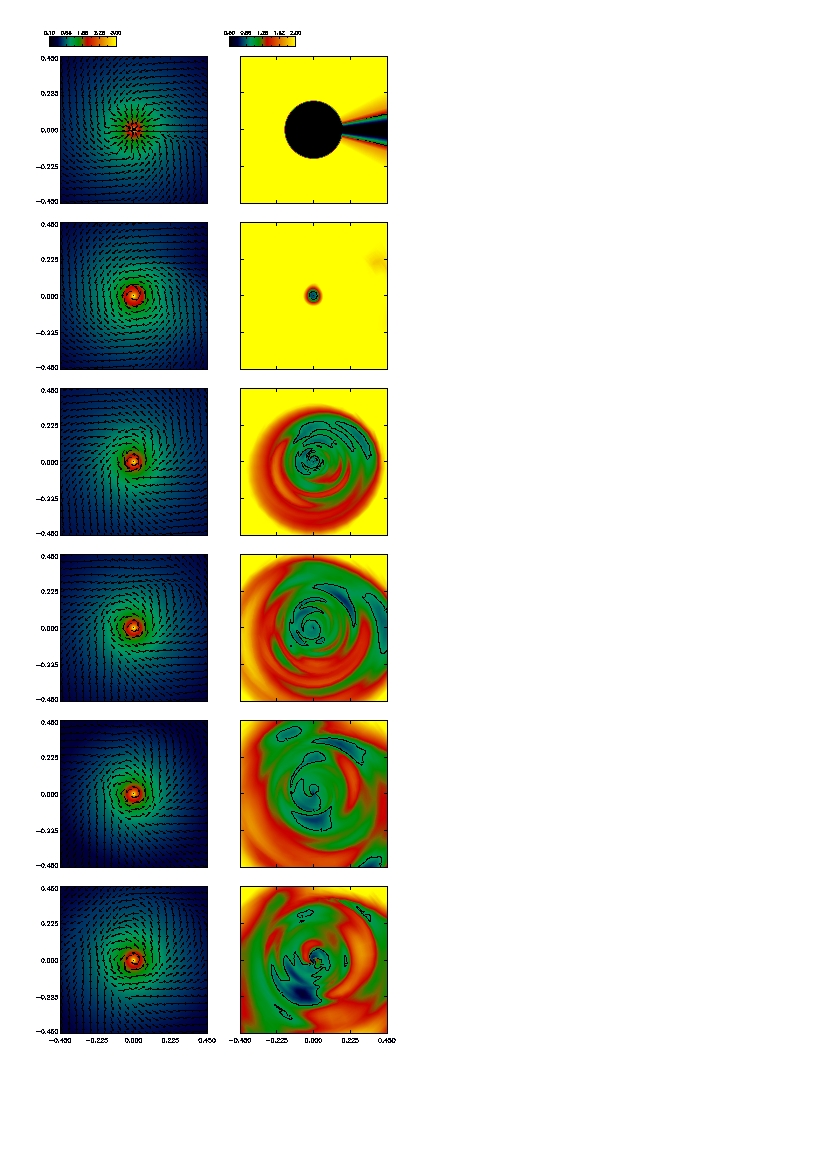}
\vspace*{-25mm}
\caption{The same as in Fig. \ref{fig:torus1} but the maps
show the equatorial plane in a zoom-out (i.e. 20 times larger scale). 
Note that the color scales
are now different than in Fig. \ref{fig:torus1}.
}
  \label{fig:torus1_large}
   \end{figure}

\begin{figure}
\hspace{-1.3cm}
\includegraphics[scale=0.7]{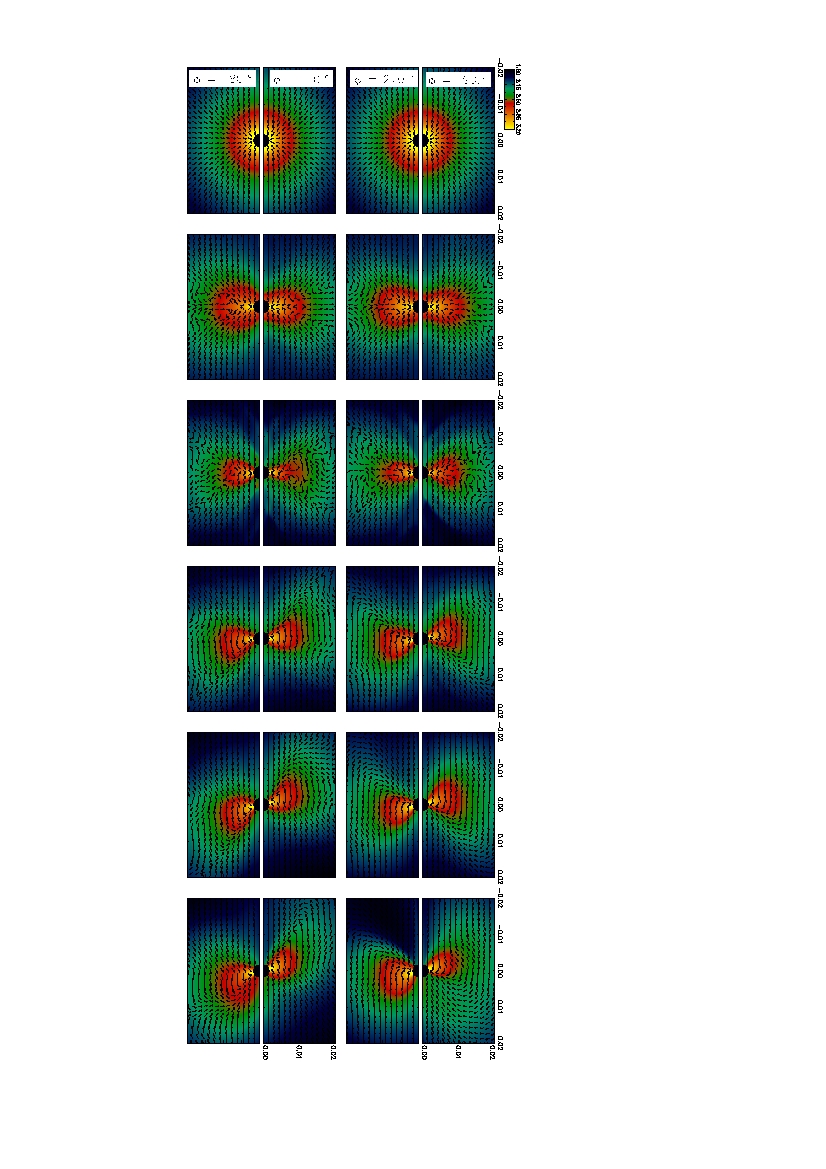}
\vspace*{-25mm}
\hspace*{-30mm}
\caption{The results
  for model $A_{32}$ (see Tab. \ref{tab:models}),
plotted in the $r-\theta$
  plane, at 4 slices of the $\phi$ angle:
  $0^{\circ}$, $90^{\circ}$, $180^{\circ}$ and $270^{\circ}$. 
The maps show density and velocity fields in the central region. 
The corresponding times are, from top to bottom: $t^{'}$ = 0.0, 0.018, 0.09, 0.16, 0.23 and 0.29. 
}
  \label{fig:torus2}
   \end{figure}

\begin{figure}
\hspace{-1.3cm}
\includegraphics[scale=0.7]{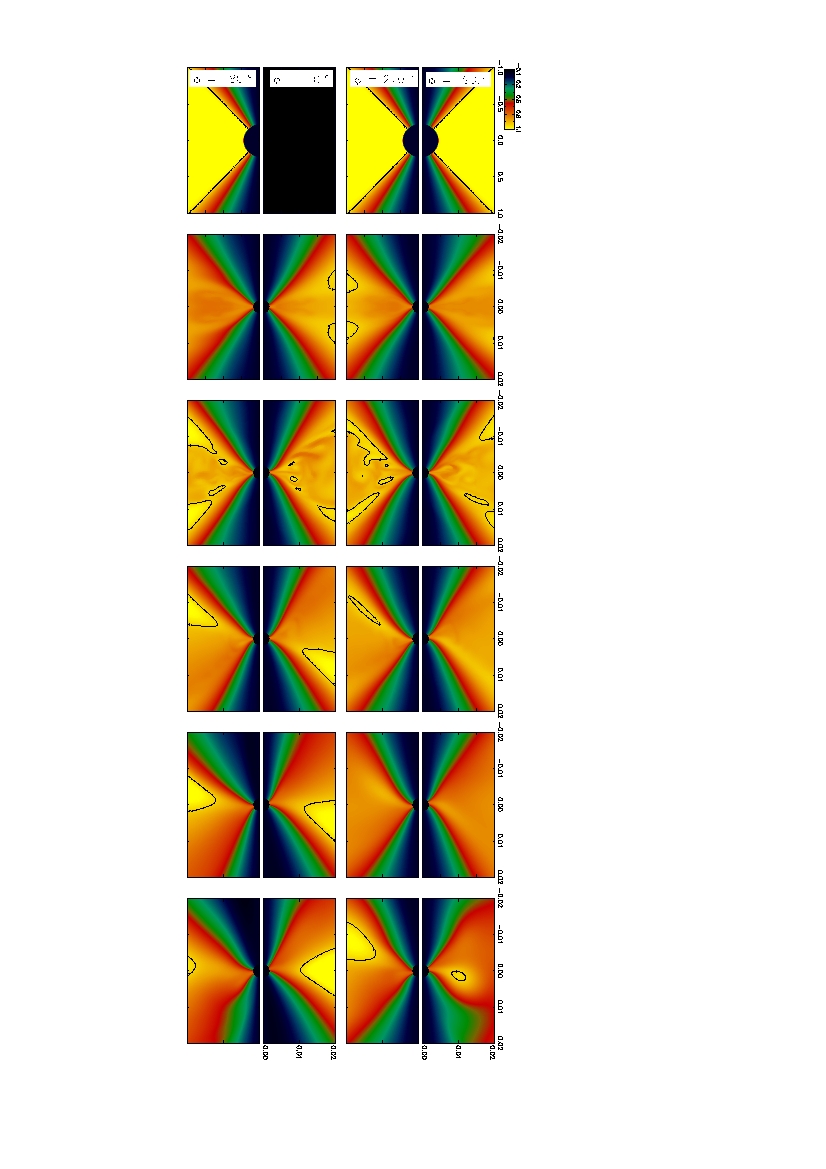}
\vspace*{-28mm}
\hspace*{-30mm}
 \caption{The maps of the specific angular momentum
  for model $A_{32}$ (see Tab. \ref{tab:models}),
plotted in the $r-\theta$
  plane, at 4 slices of the $\phi$ angle:
  $0^{\circ}$, $90^{\circ}$, $180^{\circ}$ and $270^{\circ}$, and at
$t^{'}$ = 0.0, 0.018, 0.09, 0.16, 0.23 and 0.29 (from top to bottom).
The maps show the central region (only first map is in zoomed-out).
The solid line mark the contour of $l_{\rm spec}=l_{\rm crit}$. 
}
  \label{fig:torus2lspec}
   \end{figure}

\begin{figure}
\hspace{-1.3cm}
\includegraphics[scale=0.7]{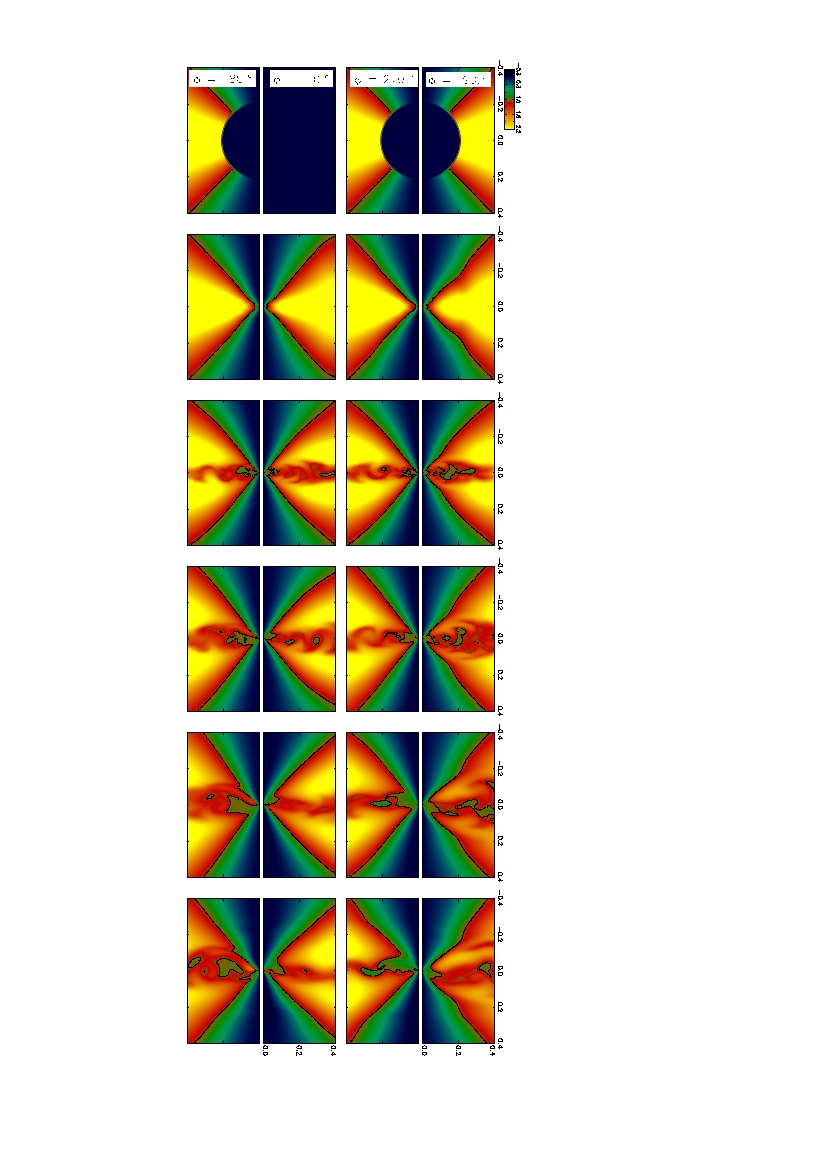}
\vspace*{-28mm}
\hspace*{-30mm}
 \caption{The same as in Fig. \ref{fig:torus2lspec} but all the maps
   are in zoomed-out. Note that the color scale is now different than in
Fig. \ref{fig:torus2lspec}.
}
  \label{fig:torus2lspec_large}
   \end{figure}

In Figures
\ref{fig:torus1}, \ref{fig:torus1_large}, \ref{fig:torus2},
\ref{fig:torus2lspec} and \ref{fig:torus2lspec_large} we show
the color coded maps of the
central region, as well as the zoomed-out. 
The maps show the density distribution and velocity field,
as well as the specific angular momentum,
 plotted  for several snapshots during
the evolution: $t^{'}=$0, 0.018, 0.09, 0.16, 0.23 and 0.29.
Note that the top-right panel in Figure \ref{fig:torus1}, as
well as all the top panels in Figure \ref{fig:torus2lspec}, i.e. for 
$l_{\rm spec}(t=0)$, are plotted on the scale 20 times larges
compared to other panels, because
in the inner region initially we assumed $l_{\rm spec}=0$.

Figure
\ref{fig:torus1} shows the density and specific angular momentum
distribution in the central region (up to 0.02 $R_{\rm B}$),
 plotted for the equatorial plane. 
The orientation of the plots
in the $x-y$ plane is standard, i.e. the $x>0$ semi-axis corresponds to 
$\phi=0$. 
The density distribution, initially spherical at $t=0$
(top left panel) changes in time, as
the material which carries specific angular momentum approaches the
center. At $t^{'}=0.018$, the gas is rotating around the $z$-axis (arrows 
over plotted on the density maps denote the direction of the 
velocity vectors with components $v_{\rm r}$ and  
$v_{\phi}$). 
The material is distributed axisymmetrically,
and the specific angular momentum is rather large (i.e. 
$0.8<l_{\rm spec}/l_{\rm crit}<1.2$)
for most of the $\phi$ directions. A clump of gas with relatively 
smaller $l_{\rm spec}$, which can be seen on the second right panel, 
 is a remaining of the gas
with $l_{\rm spec}=0$ initially present at the equator for 
 $\phi=(-15^{\circ},15^{\circ}$).
This clump is tracking the archimedean spiral 
(see also e.g. Lemaster et al. 2007), and such a trajectory
appears to be due to 
the pressure gradient force and rotation (similarly to a cyclone).

At $t^{'}=0.09$, the material with small $l_{\rm spec}$ is already mixed
with the gas of high $l_{\rm spec}$. 
The material rotates very fast in the equatorial plane,
and a circular pattern of slightly larger and smaller
specific angular momentum regions, 
visible on the third plot, form due to mixing. 
These motions are suppressed at  $t^{'}=0.16$.
For some specific directions, namely $\phi \sim 90^{\circ}$ and 
$\phi \sim 270^{\circ}$, the angular momentum near the inner radius
becomes very small (as indicated by the green spots in Fig. \ref{fig:torus1}), 
and the density in these regions drops, indicating that the
material falls radially to the black hole.

The solid line in the angular momentum maps mark the
contour at which $l_{\rm spec}/l_{\rm crit} = 1.0$.
We note that at $t^{'}=0.23$ and $t^{'}= 0.29$, in the equatorial plane 
a substantial fraction of material has  $l_{\rm spec} <
l_{\rm crit}$. This material is located at an extended region
 approximately along the diagonal of the plane
(i.e. $\phi \sim 45^{\circ}$ and $\phi \sim 225^{\circ}$, as marked by
 the orange shade in Fig. \ref{fig:torus1}).
Another region of even smaller angular momentum, $l_{\rm spec} \ll
l_{\rm crit}$, has a very limited radial extension close to the
center, and is slightly elongated along the other diagonal
(i.e. $\phi \sim 135^{\circ}$ and $\phi \sim 315^{\circ}$, as marked
by the green shade in Fig. \ref{fig:torus1}). For the
same $\phi$, but at larger distances, the specific angular
momentum in the equatorial plane is very large, and exceeds $l_{\rm crit}$.

In Figure \ref{fig:torus1_large} we show the distribution of $l_{\rm spec}$
on the larger scale. (Note also that the color scale is different 
than in Fig. \ref{fig:torus1}. The red and green shades now mark 
the regions with $l_{\rm spec}/l_{\rm crit}>1.0$).
As the Figure shows, more material with $l_{\rm spec}<l_{\rm
  crit}$ mixes in at late stages of evolution, $t^{'}=0.09$ sec.

From the Figs. \ref{fig:torus1} and  \ref{fig:torus1_large}, 
we conclude that a torus starts forming in the innermost regions in the 
equatorial plane at $t^{'}=0.018$,
and it is 
rotationally supported for all the $\phi$ directions, because the low
$l_{\rm spec}$ material is quickly mixed in.
At  $t^{'}=0.09$,
the distribution of density and specific angular
momentum is almost axisymmetric, but only in the inner region.
However as the zoom-out maps show, the gas with $l_{\rm spec}< l_{\rm
  crit}$ is mixing in and is distributed asymmetrically.
The innermost symmetric configuration lasts until about 
 $t^{'}=0.16$,
and after this time the 
torus position and shape changes. To investigate what really
happens,
we need to look at the flow from a different perspective.
Therefore, in Figures \ref{fig:torus2},  \ref{fig:torus2lspec} 
and \ref{fig:torus2lspec_large}, we show the
slices perpendicular to the equatorial plane.
 The maps show density and velocity field,
as well as the specific angular momentum,
as seen from $\phi=0^{\circ}$, $90^{\circ}$, $180^{\circ}$ and
$270^{\circ}$, at the same times as in Figs.\ref{fig:torus1} and  \ref{fig:torus1_large}.

As shown in the Figure \ref{fig:torus2},
at 
 $t^{'}=0.018$,
the material is indeed accumulated near the equator,
while at the poles the density is much lower. The axial symmetry is
not perfect, since at $\phi=180^{\circ}$ the torus is
geometrically thicker than at other $\phi$ directions, 
which corresponds to the location of the 'clump'
with smaller specific angular momentum (c.f., Fig. \ref{fig:torus1}).
As indicated by the arrows (velocity vectors with $v_{\rm r}$ and $v_{\theta}$ 
components), an equatorial outflow occurs at most of the $\phi$
directions, however at $\phi=180^{\circ}$ this outflow is weaker,
because of slower rotation.

At 
 $t^{'}=0.09$,
the torus is relatively thin and located at the equator 
in every $\phi$ direction, while at the poles the density is very low.
The gas accretes onto
the center mostly through the poles, while at the equator the
flow pattern is complex (circulations, outflows).
The flow complexity is reduced at 
$t^{'}=0.16$,
because the gas that flows into the equatorial region has less angular
momentum and can directly accrete onto the black hole,
finding its way along one of the poles.
At $\phi=0^{\circ}$ and $\phi=270^{\circ}$ the gas turns towards the
northern pole (i.e. $+z$), while at $\phi=180^{\circ}$ and $\phi=90^{\circ}$
the flow turns towards the southern pole (i.e. $-z$).
A line which would mark the regions of the maximum
density, is now tilted with respect to the equator by an angle of $\sim
20^{\circ}$.
This means that the
torus which at 
$t^{'}=0.09$,
was almost symmetric with respect to 
the equatorial plane, is tilted after 
$t^{'}=0.16$.

In Figure \ref{fig:torus2lspec} we show the 
maps of specific angular momentum, 
also perpendicularly to the equatorial plane and
for the same $\phi$ directions as in Fig. \ref{fig:torus2}.
As the Figure shows, the specific angular momentum distribution
 is symmetric with respect to the
equatorial plane at 
$t^{'}=0.09$,
and nearly symmetric at 
$t^{'}=0.16$.
The axial symmetry is not perfect, however the regions of large
angular 
momentum ($l_{\rm spec} \ge l_{\rm crit}$)
appear in every slice.
At 
$t^{'} \ge 0.16$,
the flow is no longer symmetric, and the regions with very
large  specific angular momentum
appear either below the equator (at $\phi=0^{\circ}$ and $\phi=90^{\circ}$)
or above it (at $\phi=180^{\circ}$ and $\phi=270^{\circ}$).
These regions correspond to a flatter torus, i.e. relatively
thin on one side, while the gas which is not rotating fast makes the
configuration geometrically thicker.
We notice, that the lack of the top-bottom symmetry in density and
$l_{\rm spec}$ maps at this phase of
system evolution is a consequence of the earlier non-axisymmetry.
This asymmetry was introduced to the velocity field in the initial
conditions at large radii, and subsequently propagated to the inner
radii and affected the density distribution there, as soon as the
rotating gas reached there.

In Figure \ref{fig:torus2lspec_large} we show the distribution of the
specific angular momentum in the zoom out.
The flow is rotationally supported in the outer
regions. The $l_{\rm spec}$ distribution in the flow is asymmetric 
 at large scales (there is significantly more material with
large $l_{\rm spec}$  for the directions of $\phi=0^{\circ}$ and
$270^{\circ}$ 
than for $\phi=90^{\circ}$ and $180^{\circ}$). However, this is the
case only in the
inner region, i.e. the torus is tilted, while
outer regions, even at late times, are rather symmetric 
with respect to the equatorial plane.

\begin{figure}
\includegraphics[scale=0.65]{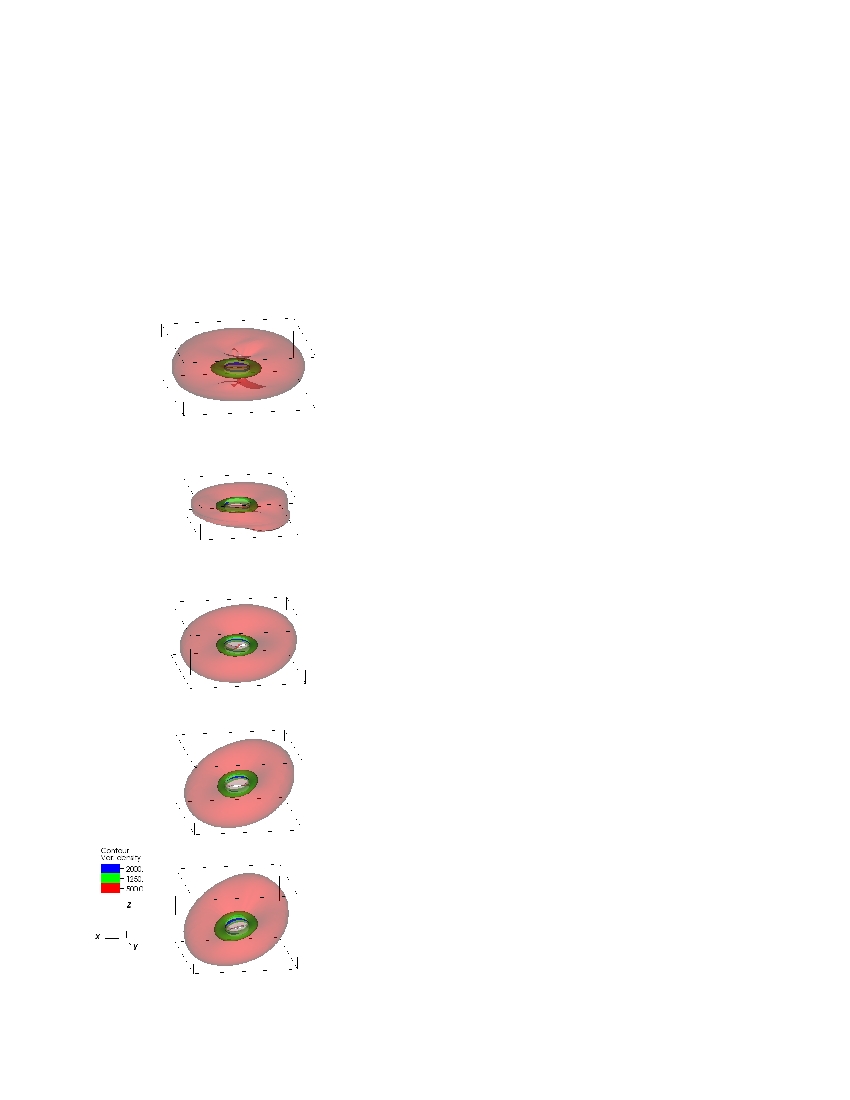}
\vspace*{-20mm}
\caption{Density isosurfaces for five time snapshots: 
$t^{'}$ = 0.018, 0.09, 0.16, 0.23 and 0.29,
  for model $A_{32}$ (see Tab. \ref{tab:models}). The colors mark the surfaces of 
$\rho = 500$, 1250 and 2000 $\rho_{\infty}$. The boxes are
  scaled to the density isocontour $\rho=500$ $\rho_{\infty}$ and show the 
central region of the radius about 0.004 $R_{\rm B}$.
}
  \label{fig:density3d330}
   \end{figure}

To visualize the 3-D configuration  better,
in Figure \ref{fig:density3d330} we show the density isosurfaces in 3-D. 
The plots show 3 arbitrarily chosen contours of the constant density:  
2000, 1250 and 500 
$\rho_{\infty}$,
which correspond to the gas densities
very close to the inner radius, i.e. inside $\sim 0.004 R_{\rm B}$. 
The maps are plotted for 5 different time snapshots: 
$t^{'}=0.018$, 0.09, 0.16, 0.23 and 0.29.
We do not show the density distribution at t=0, because it is
purely spherical. The orientation of the figures is almost edge-on,
i.e. the
$z$-axis is the rotation axis of the system, and $x-y$ plane is the
equatorial plane.

These density contours may be regarded as the shapes of the torus
(however one should keep in mind that the material with smaller/larger
density is also present there). Therefore, as the Figure shows, the
torus
is closed already at 
$t^{'}=0.018$.
After 
$t^{'}=0.16$,
the configuration becomes
tilted with respect to the equatorial plane, and at 
$t^{'}=0.29$
this tilt is
the largest. After 
$t^{'}=0.29$
the torus starts precessing, and the
 precession period is very long
(by the end of the run, the ring precessed around the $z$-axis 
 by less than $90^{\circ}$).
The torus precession was also shown in the right panel of the Figure 
\ref{fig:levol} (Sec. \ref{sec:reference}). The changing
values of $L_{\rm x}$, $L_{\rm y}$ and $L_{\rm z}$, show that at 
$t^{'} \sim 0.1$
the rotation axis, which is defined by the direction of {\bf L},
 starts to tilt towards the first quarter of the
$x-y$ plane, and after 
$t^{'} \sim 0.29$
the rotation 
axis moves counter-clockwise.

\begin{figure}
\includegraphics[scale=0.65]{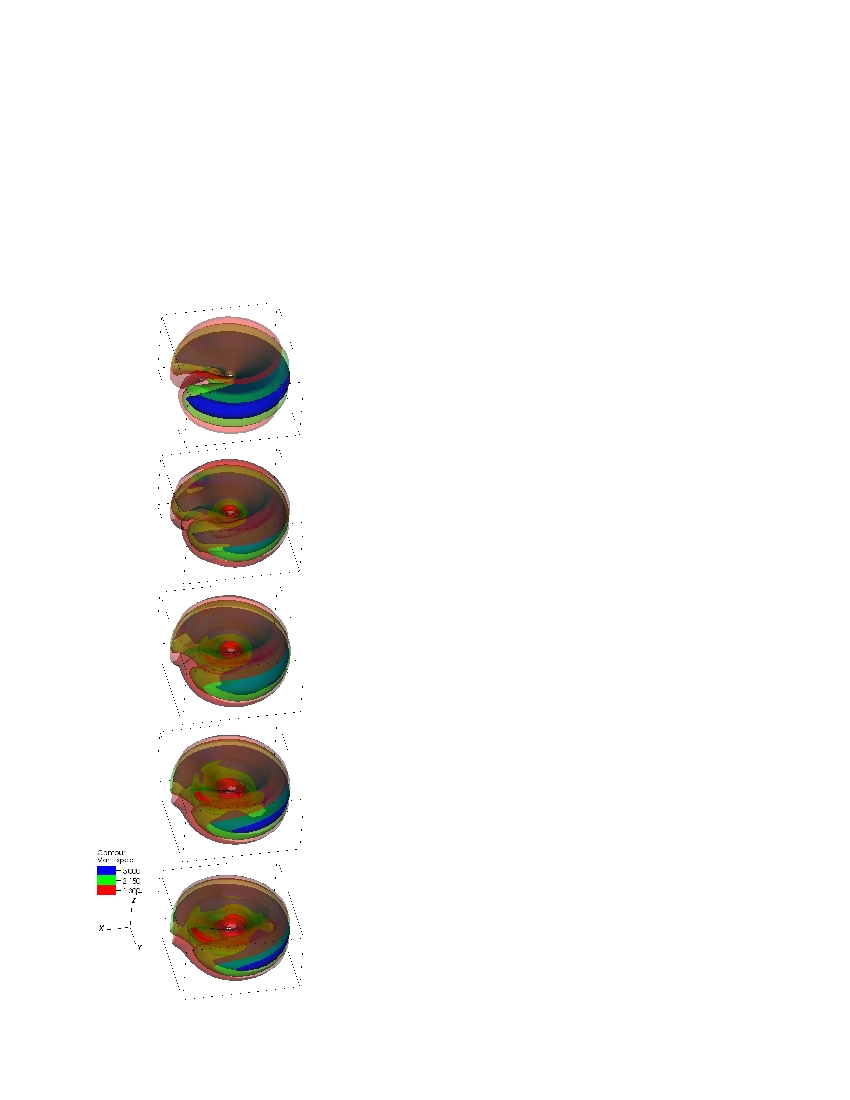}
\vspace*{-20mm}
\caption{Isosurfaces of the specific angular momentum in five time snapshots: 
$t^{'}$ = 0.018, 0.09, 0.16, 0.23 and 0.29,
  for model $A_{32}$ (see Tab. \ref{tab:models}). The colors mark the
  surfaces of  $l_{\rm spec}/l_{\rm crit}=1.3$, 2.15 and 3.0. The boxes are
  scaled to the  isocontour of $l_{\rm spec}/l_{\rm crit}=1.3$ and show the 
region of the radius about 0.6 $R_{\rm B}$.
}
  \label{fig:lspec3d330}
   \end{figure}

In Figure \ref{fig:lspec3d330} we show the isosurfaces of the specific
angular momentum in 3-D. 
The contours are for $l/l_{\rm crit} = 1.3$, 2.15 and 3.0, and
correspond to the zoomed-out regions shown in
Fig. \ref{fig:torus2lspec_large}
(the radial extension of about 0.6 $R_{\rm B}$).
As the Figure shows, the material with various angular momentum is
mixing in the innermost region at 
$t^{'} = 0.018$
while at the
outer parts initially the momentum is not mixed. At later times, the 
outer parts of the flow contain more and more mixed angular momentum
layers, while in the innermost region the gas rotates slower than
at the beginning, and the angular momentum distribution is smoother.

\begin{figure}
\includegraphics[scale=0.7]{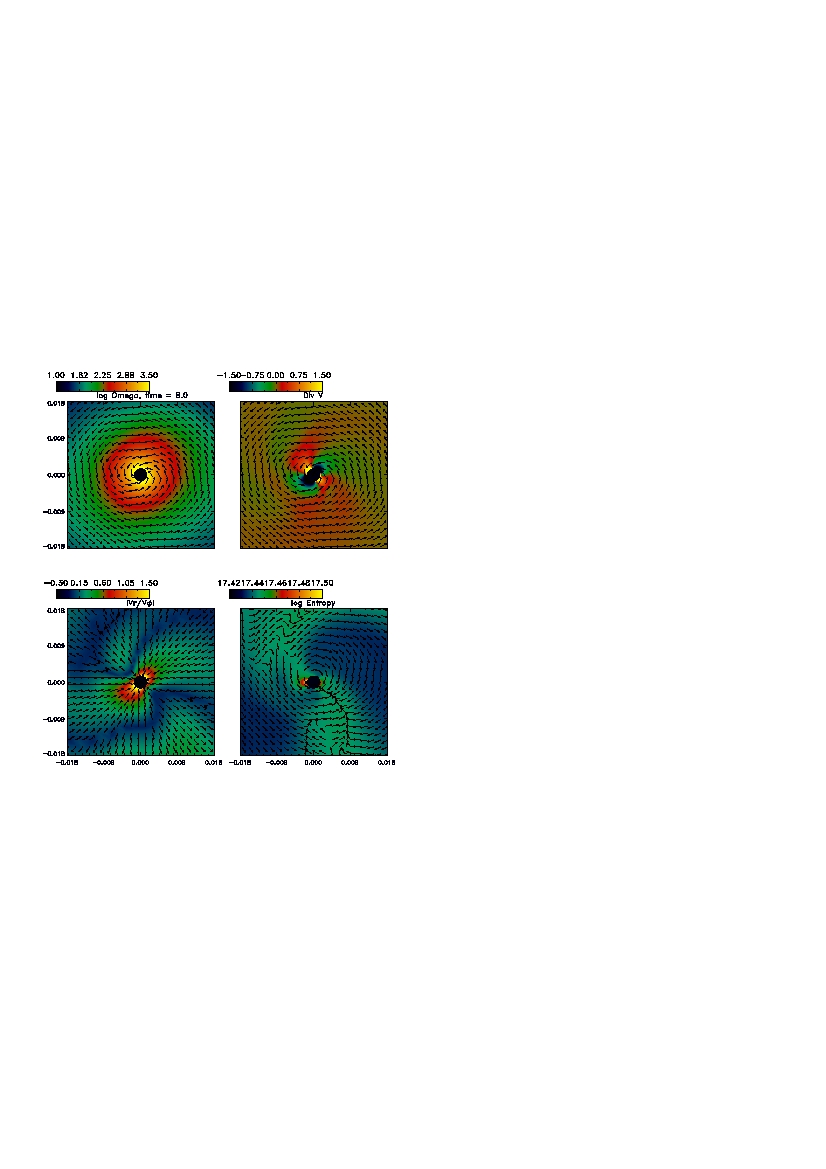}
\vspace*{-70mm}
 \caption{The results
  for model $A_{32}$ (see Tab. \ref{tab:models})
plotted in the equatorial  plane, at $t^{'}$ = 0.29. 
 The maps show:  angular velocity (upper left),  velocity divergence 
(upper right), radial to azimuthal velocity
  ratio (bottom left) and  entropy (bottom right). The arrows denote 
the directions of velocity vectors with $v_{r}$ and $v_{\phi}$ components (upper two
  panels), with only  $v_{r}$ component (bottom left panel)
or vorticity vectors with $w_{r}$ and $w_{\phi}$ components (bottom right
  panel).
}
  \label{fig:torus3}
   \end{figure}

In Figure \ref{fig:torus3} we show the maps
of entropy, $S$,  radial to azimuthal velocity ratio, $v_{\rm r}/v_{\phi}$,
angular velocity, $\Omega$, and velocity divergence, ${\bf div ~v}$,
as
calculated close to the end of this simulation, at time 
$t^{'} = 0.29$.
The maps show the inner region in the equatorial plane.
Close to the center, the angular velocity is the largest along the diagonal
of the plane ($\phi \sim 45^{\circ}$ and $\phi \sim
225^{\circ}$), which corresponds to the cross-section line along which
the torus crosses through the equatorial plane (cf. Fig. \ref{fig:density3d330}).
This line also corresponds to the largest entropy, as well as positive
velocity divergence, while at the other diagonal there is smaller entropy
and negative velocity divergence.
This means that the fluid is compressible (and supersonic).

Our analysis of
the 3-D results shows
the flow has a largest negative divergence close to the poles
while  the largest positive divergence 
is somewhat above and below the
equator, i.e. on the surface of the torus,
as well as at its cusp. 
From the poles, the gas flows radially onto the center with large
supersonic velocities, i.e. $M_{\rm r} = \sqrt{v_{\rm r}^{2}/c_{\rm s}^{2}} \gg 1$. Close to the equator,
the flow is captured in the torus and the radial velocities are
smaller,
so $M_{\rm r}< 1$. Still, the gas rotates very fast, and the 
total Mach number is large, $M_{\rm tot} \gg 1$, because of the
contribution from the azimuthal velocity.

The entropy in the flow should be constant along the streamlines, however may
vary from one streamline to the other. The entropy 
gradient corresponds to a non-zero vorticity in the flow.
The direction of vorticity arrows (${\bf w} = rot \times {\bf v}$) 
in the bottom-right panel indicate
that the torus is rotating counter-clockwise. 
The direction of radial
velocity arrows in the bottom-left panel confirms that 
the radial inflow occurs from the directions of the smallest angular velocity.

The above considerations led us to the conclusion that qualitatively, 
the basic pattern of the torus evolution
is uniform. It does only weakly depend on whether we 
assume the axisymmetric initial conditions, 
or if the initial distribution of angular momentum was
perturbed (provided that this perturbation was
small enough to allow for the torus formation).
The axisymmetry was explicitly assumed by PB03 throughout their
2-D simulations, as well as in our 3-D reference models in the initial
 conditions.
The similarity of the torus behavior that we find here means that in
both 3-D models $R$ and in the $A$ models, the rotationally
supported tori form after time 
$t^{'} = 0.018$,
than they exhibit strong
equatorial outflows (which stabilize them), 
then they become tilted with respect to the equatorial plane due to the 
asymmetric polar accretion,
 and finally start precessing. What differs in the models,
is the moment when the outflow stops and when the torus becomes
tilted, as well as the tilt angle (it is about twice as large 
in model $A$ 
than
in model $R$
at the end of our simulations;
see also Fig. \ref{fig:levol}).
These two features (i.e. tilt and precession) 
cannot be
 studied in the 2-D simulations,
but are detected in the 3-D models.

However, we note that the similarity between the $R$ and $A$
models is less pronounced when considering the details, e.g., of the
torus shape. The asymmetric 
torus can be thicker (warped), or thinner, depending on $\phi$.
To investigate further the difference between $R$ and $A$ models and
the role of non-axisymmetry, we calculated the angular momentum 
at the inner boundary, as a function of the angles 
$\theta$ and $\phi$.

\begin{figure}
\plotone{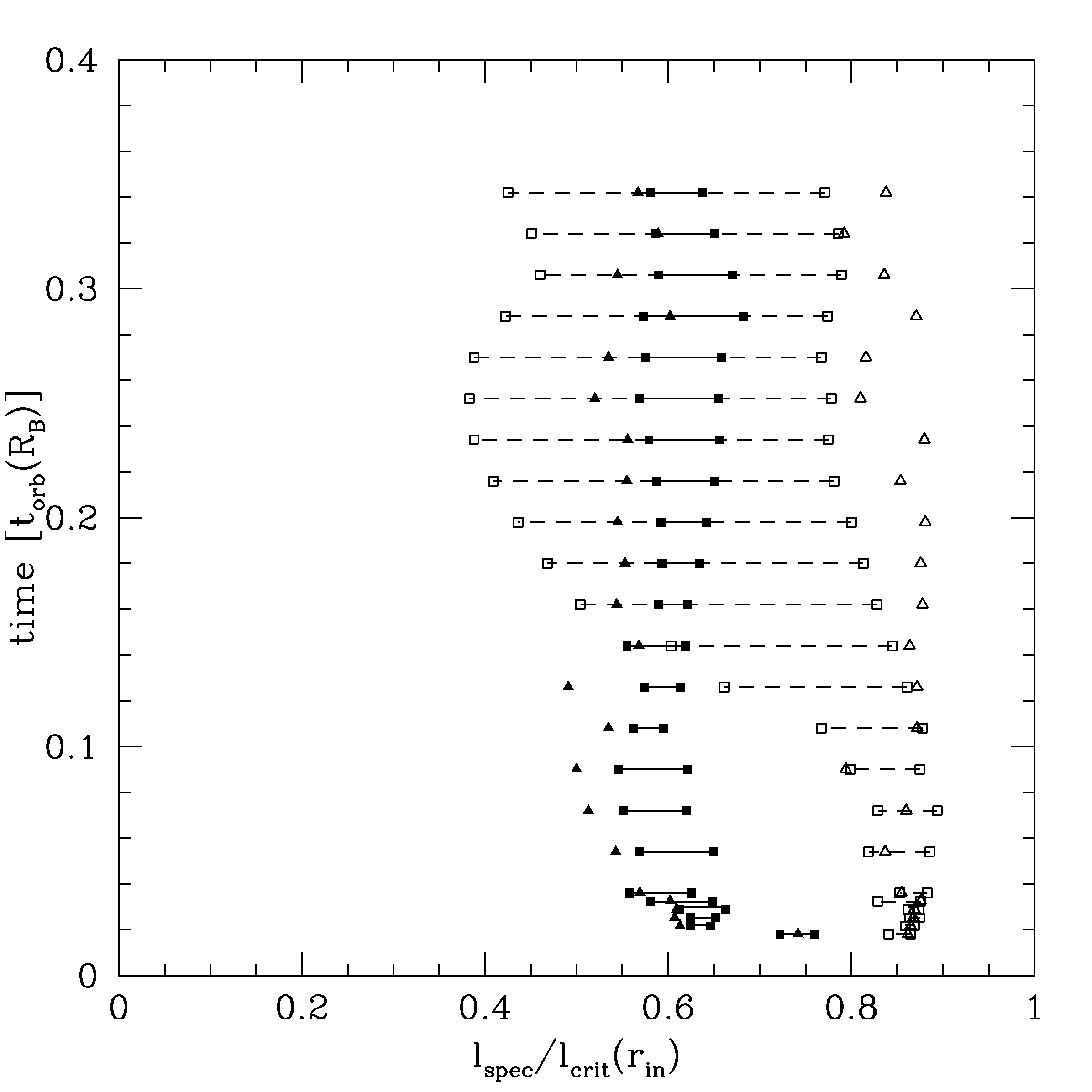}
\caption{The time evolution of the 
specific angular momentum at the inner radius. The results are shown
for the  axisymmetric reference model $R$ (triangles) 
and for the model $A_{\rm 32}$ (squares, showing the range of values
for different $\phi$-directions). The  filled symbols and
solid lines denote the $\theta$-averaged results, and  the open
symbols are for at the equatorial plane.}
  \label{fig:phiscatter}
   \end{figure}

In the axisymmetric model $R$, the specific angular momentum at
$r_{\rm in}$
depends only on $\theta$ by definition. After the torus is formed, the maximum
value at the equator is $l_{\rm spec}/l_{\rm crit}(r=r_{\rm in},\theta=90^{\circ})\approx
0.85$,
while the value averaged over the angle $\theta$ is  
$\bar l(r=r_{\rm in})\approx 0.55$.
For the non-axisymmetric case (in particular, model $A$), the results
depend also on the angle $\phi$. Therefore both the equatorial angular
momentum, $l(r=r_{\rm in},\theta=90^{\circ})$, and the
$\theta$-averaged, $\bar l(r=r_{\rm in})$, are scattered and do not have
to match with the  axisymmetric solution. 
In Figure
\ref{fig:phiscatter}, the results for the axisymmetric
model, are denoted by single points, while
the non-axisymmetric solutions 
are represented by the horizontal lines to show the scatter in $\phi$.

The level of non-axisymmetry is represented as the spread between the
maximum and minimum values of $l_{\rm spec}$ at a given time.
For the model $A$, in the beginning of the torus
evolution,
i.e 
$0.018 < t^{'} < 0.09$,
the equatorial and averaged values of
$l_{\rm spec}$ match well with the axisymmetric solutions and are only
slightly
 scattered with $\phi$. The equatorial value is much larger than
 the $\theta$-averaged, which means that the torus is located in the equatorial
 plane where $l_{\rm spec}$ is the largest.
As the evolution proceeds, 
$t^{'} \ge 0.09$,
the scatter with $\phi$ increases and the range of
$l_{\rm spec}$ does not match the axially symmetric
solution. However, the equatorial angular momentum is still much
larger than the average. This means that the  torus is
located at the equatorial plane, but it is now asymmetric.
After 
$t^{'} \ge 0.16$,
the situation changes, and $l_{\rm spec}$ at the
equator has a very large scatter, being either smaller or larger than
the $\theta$-averaged value (i.e. the solid and dashed
lines overlap in the Figure). 
The averaged angular momentum has also some scatter,
but much smaller than the equatorial one. 
This indicates that the torus is not symmetric, and is tilted with respect
to the equatorial plane.

\begin{figure}
\voffset=-8 cm
\includegraphics[scale=0.7]{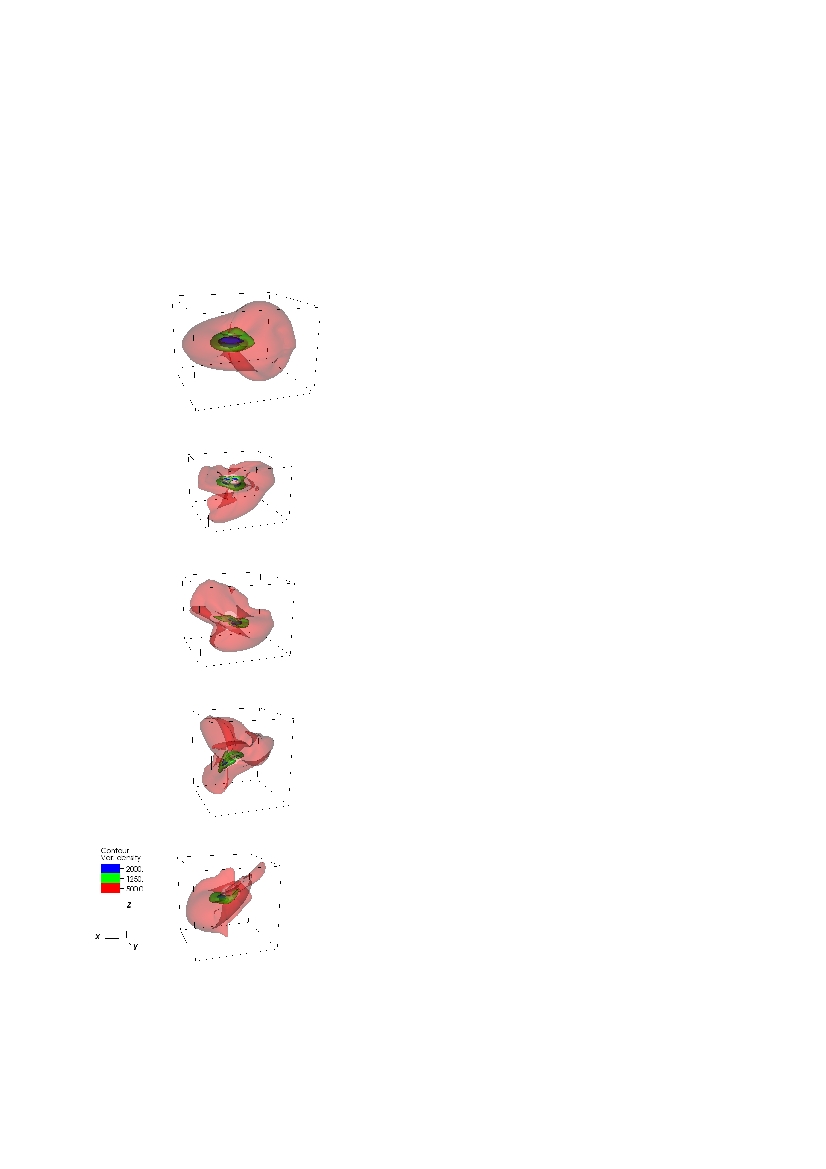}
\vspace*{-40mm}
\caption{ Density isosurfaces for five time snapshots: 
 0.018, 0.09, 0.16, 0.23 and 0.29 $t_{\rm orb}(R_{\rm B})$,
  for one of the 'broken torus' models,
 $B_{60}$ (see Tab. \ref{tab:models}). The colors mark the surfaces of 
$\rho = 500$, 1250 and 2000 $\rho_{\infty}$. The boxes are
  scaled to the density isocontour $\rho=500$ $\rho_{\infty}$ and show the 
central region of the radius about 0.004 $R_{\rm B}$.
}
 \label{fig:fi240}
   \end{figure}

We performed similar analysis of
 other non-axisymmetric models, $B$-$E$
(cf. Table \ref{tab:models}).
 For models $B$ and $C$ ($\Delta \phi_{0}=240^{\circ}$ and $\Delta \phi_{0}=120^{\circ}$),
 at the beginning of the simulation 
$0.018 < t^{'} < 0.07$,
 the equatorial value of $l_{\rm spec}$ is always larger than the averaged, 
 and the scatter with $\phi$ is smaller in model $B$ than that in model $C$.
 This indicates that a rotationally supported torus is present in the 
equatorial plane.
 At later times, the equatorial $l_{\rm spec}$ becomes equal or smaller 
than average,
 and the scatter in both models $B$ and $C$ is 
quite substantial (larger than in model $A$). 
This implies
 that a torus, which possibly tried to form  at the early phase,
 is broken (i.e. not completely closed), 
as well as tilted from the equator.
An example of such a 'broken torus' configuration is shown in Figure
\ref{fig:fi240}.

For models $D$ and $E$ 
($\Delta \phi_{0}=30^{\circ}$ and $\Delta \phi_{0}=60^{\circ}$),
the scatter in both $\theta$-averaged and equatorial $l_{\rm spec}$ is
very large and does not decrease with time 
(up to $t^{'} \sim 0.018$).
The
averaged $l_{\rm spec}$ can be larger or smaller than the axisymmetric one,
while the equatorial $l_{\rm spec}$ in these models is always smaller 
than that in the
axisymmetric case, and locally (i.e. for some $\phi$ angles) can be 
smaller than the $\theta$- averaged. 
 We conclude that in these models the torus does not form, the
 solution is not axisymmetric, and
 the gas with very small angular momentum can accrete onto the black hole 
 through the equator. 

We note that in this sense the properties of these models
are similar to those of the model $A$ in later times. However,
when comparing the density distributions, in the models  $D$ and $E$
the gas is always distributed much more uniformly, i.e. it does not
concentrate neither close to the equator nor to any specific plane. 
In model $A$, the gas density near the poles is
always orders of magnitude lower than that at the equatorial plane, or the
plane tilted to the equator by a small angle.

This is not the case for models $D$ and $E$, for which the torus does
not form.
For models $B$ and $C$, the gas is concentrating near the equator, 
but only for some range of $\phi$-directions, i.e. the torus is not closed.

\begin{figure}
\plotone{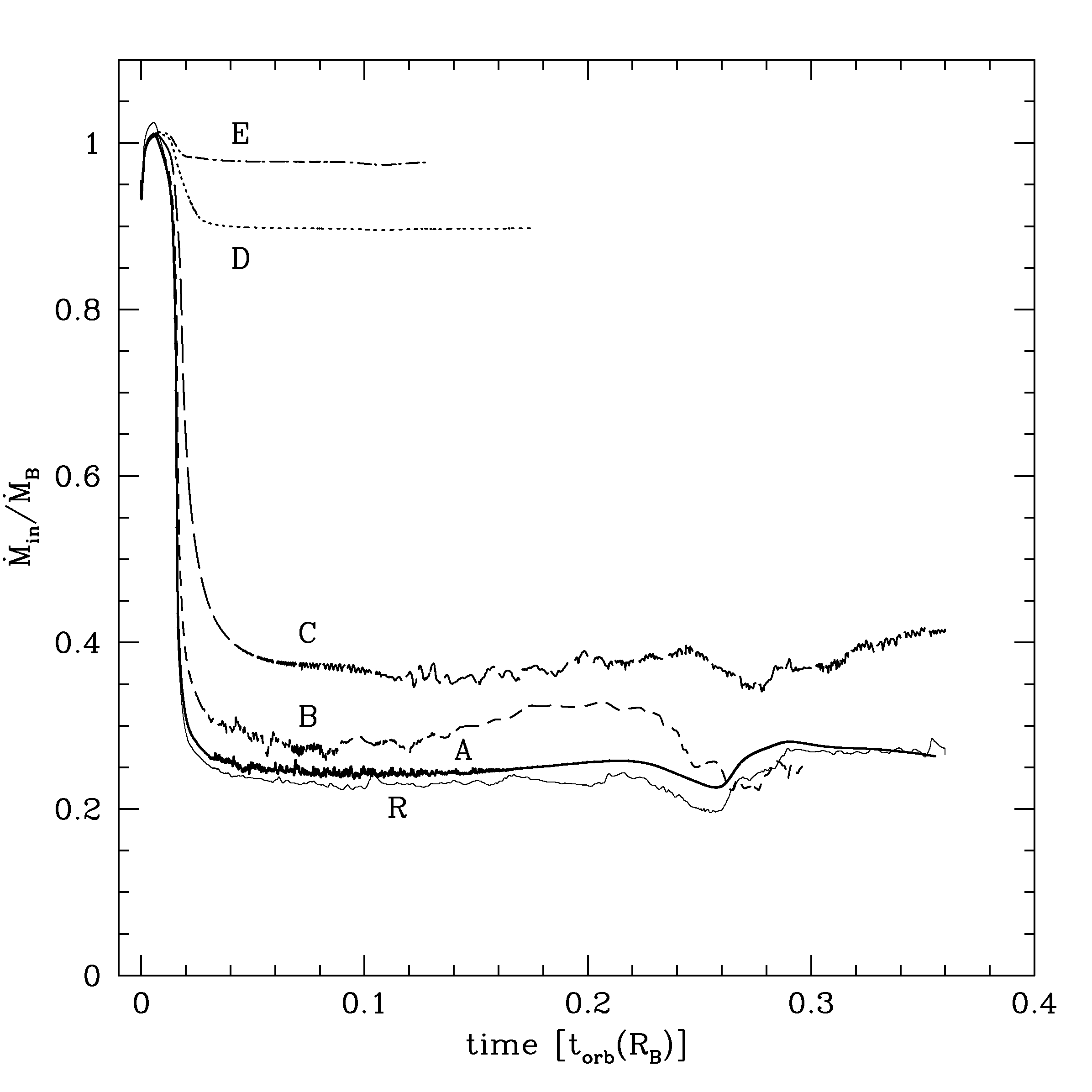}
\caption{Time evolution of the mass accretion rate
  through the inner
  boundary ($\dot M_{\rm in}$), in units of the
  Bondi rate ($\dot M_{\rm B}$). The non axisymmetric models are: 
$A$ (thick solid line),
$B$ (short dashed line), $C$ (long dashed line),
$D$ (dotted line), $E$ (dot-dashed line) and reference
axisymmetric model $R$ (thin solid line).
}
  \label{fig:Mintime}
   \end{figure}

Figure \ref{fig:Mintime} shows the time evolution of the
accretion rate through the inner boundary, $\dot M_{\rm in}$ 
for the non-axisymmetric models.
Before the rotating material approaches the black hole 
($t^{'} < 0.018$),
$\dot M_{\rm in}$
 is equal to the Bondi accretion
rate for all models. Once the gas starts rotating also in the innermost
parts, the accretion rate drops
reaching about 25\% - 40\% of the Bondi rate for models $A$, $B$, $C$
and $R$.
This is because much more material gets captured in the rotating torus, and
does not fall radially into the black hole.
The model $A$ 
 gives the lowest accretion
rate with rather small and regular variability pattern, very close to that
obtained in the reference model $R$.

\begin{figure}
\plotone{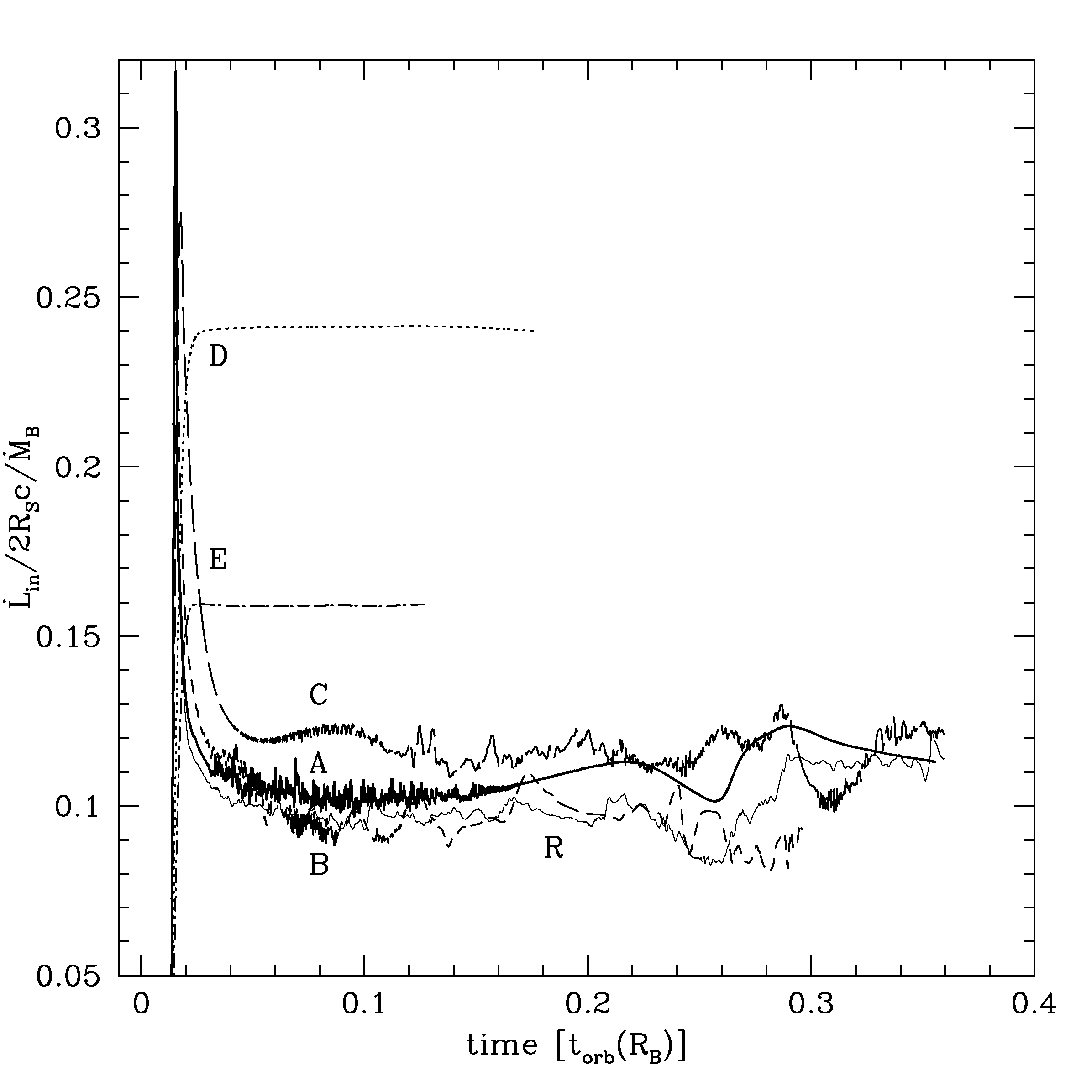}
\caption{Time evolution of the angular momentum flux through the inner
  boundary, in units of the critical angular momentum times the Bondi
  accretion rate. 
The non axisymmetric models are: $A$ (thick solid line),
$B$ (short dashed line), $C$ (long dashed line),
$D$ (dotted line), $E$ (dot-dashed line) and the reference
axisymmetric model $R$ (thin solid line).
}
  \label{fig:Lintime}
   \end{figure}

Figure \ref{fig:Lintime} presents the evolution of the angular momentum flux through the inner
  boundary: $\dot L_{\rm in} = \int l_{\rm spec} \rho v_{r} ds$, in units of the
  critical angular momentum $l_{\rm crit}$ 
and renormalized by the
  value of the Bondi accretion rate.
At the beginning of the simulation, $\dot L_{\rm in} = 0$, since there is no
  rotation in the vicinity of the black hole. Once the
  rotating matter reaches the inner boundary, the angular momentum
starts accreting to the center.
When the torus starts forming, the fast
  rotation near the center leads to a fast rise in $\dot L_{\rm in}$. 
  However, after  several orbital cycles the outflow begins and the net radial
  velocity drops, as well as drops the density near the polar regions,
 therefore the $\dot L_{\rm in}$ is rather small
  during the torus evolution. 

Note that the quantity plotted in the Figure \ref{fig:Lintime} is not
  a flux of specific angular momentum, but the total one.
This corresponds to the amount of angular momentum which may be 
 transferred to the black hole and used to spin it up.
However, as the Figure shows, the total angular momentum  which the
  black hole could gain during our simulation, is extremely low:
$a = (cJ)/GM^{2} \approx 4\times10^{-6} - 10^{-5}$, where 
$J=\dot L_{\rm in} \Delta t$.

Figure \ref{fig:Lintime} shows
two trends in the magnitude of $\dot L_{\rm in}$.
For small $\Delta \phi_{0}$ (models $E$ and $D$), the
  rotation at inner boundary is very small, but the density is high
both in the equator and in the polar regions (still close to the spherical
  accretion). Therefore, in model $D$, with faster rotation,
  $\dot L_{\rm in}$ is larger.
For large  $\Delta \phi_{0}$ (models $A$, $B$ and $C$), the material
  accumulates rather close to the equator, at least for some $\phi$
  angles, while at the poles the density is small. Therefore regardless of
  the fast rotation, these models give systematically smaller $\dot
  L_{\rm in}$ than those in the models $E$ and $D$. 
Also, this is why the
  model $B$ gives smaller $\dot  L_{\rm in}$ than that in the model $C$.
Model $A$ is the only one in which the torus is closed (the gas
  rotates fast at the equator at every $\phi$
  angle). This leads to a larger $\dot  L_{\rm in}$ than that in model $B$,
  which is again less affected by the density distribution.

In the models $A$, $B$ and $C$, the accretion rate and $L_{\rm in}$
are variable.
For model $A$,
 a characteristic wave pattern can be seen in the
specific angular momentum distribution in the equatorial 
plane (see Fig. \ref{fig:torus1} at  
$t^{'} \sim 0.09$).
This behavior is reflected in the variable
$\dot M_{\rm in}$ and $\dot L_{\rm in}$, due to the variable radial
velocity and nearly constant density at the inner boundary. 
At time 
$t^{'} \sim 0.16$,
when the accretion rate stops
varying rapidly, the corresponding 'waves'
 in specific angular momentum map are smoothed out.
The outflow of the gas  at this time is suppressed, and the gas
accretes onto the center through the poles. As a consequence, the mass flux and
angular momentum flux through the inner boundary slightly increase,
and the curves are smoother.
The subsequent drop of both of these quantities at time 
$t^{'} = 0.25$
is
caused  by the density decrease at the inner edge, when the torus
is tilted with respect to the equatorial plane. The density at inner radius
increases again at 
$t^{'} = 0.29$
and the torus starts precessing.

The behavior of the flow in models $B$ and $C$ 
is more chaotic. The accretion rate and $\dot L_{\rm in}$ vary in time 
until the end of our simulations. Also, no precession was detected,
because in principle it would be hard to determine the boundaries of
the structure which might be precessing. We cannot use a density
threshold to define a torus, because such a torus is not a closed ring
in these two models.

\section{Discussion}
\label{sec:diss}

This paper presented the fourth phase of our study of slightly rotating 
accretion flows onto  black holes (see PB03, Proga \& Begelman 2003b, and
Proga 2005 for the first, second and third phase). 
Here we followed PB03, but we considered
3-D not 2-D axisymmetric effects. As in PB03 we made a few simplifications.
For example, we neglected  the gravitational field due to the host galaxy, 
radiative heating and cooling, viscosity and MHD effects.
Perhaps the most important 
simplification we made is neglecting the transport of energy and angular 
momentum outward as needed to accrete matter with a specific angular momentum 
higher than $l_{\rm crit}$. As shown Proga \& Begelman (2003b) for
the axisymmetric case, magnetic fields, that can drive  the transport, 
can dramatically alter the flow solution.
Still our HD results  provide a useful exploratory
study of accretion onto black holes as they have revealed unexpected 
properties and complexity of accretion flows even with simplified physics.
To our best knowledge there have not been any MHD simulations
of rotating flows where a closed torus was either assumed or failed to form.
Our results show that such flows are plausible
(see also Loeb 2004) and motivate new simulations.
In what follows we summarize and discuss our results.

We have performed numerical 3-D  hydrodynamical simulations
of slightly rotating, inviscid accretion flows onto a black hole. 
As in PB03, we attempt 
to mimic the boundary conditions of classic Bondi accretion flows with
the only modifications being the introduction of a small, latitude-dependent 
and also azimuth-dependent 
angular momentum at the outer boundary and a pseudo-Newtonian gravitational 
potential. The adopted form of the distribution of $l_{\rm spec}$ 
the density distribution at infinity to approach spherical symmetry,
because the centrifugal force is negligible. 

For the latitude-dependent angular momentum, 3-D simulations confirm 
axisymmetric results. Namely, the material that has too much angular momentum 
to be accreted forms a thick torus near the equator. 
Therefore the geometry of the polar funnel, where 
material is accreted, and the mass accretion rate through it are constrained 
by the size and shape  of the torus {but by  the outer conditions}. 
However, in 3-D the torus precesses and
is non-axisymmetric  even for axisymmetric conditions
at large radii. For the latitude- and azimuth-dependent angular
momentum in the initial conditions,
the non-rotating gas near the equator can significantly
affect the evolution of the rotating  gas. It can prevent 
closing, in the azimuthal direction, of
the rotating gas and the proper torus does not form. In such cases,
the mass accretion rate is only slightly less than Bondi rates.

Simulations with none or a small amount of a non-rotating gas
near the equator show that a torus forms and 
limits the accretion rate. However, a non-rotating gas
near the equator can inhibit torus formation and the accretion
rate will be close to the Bondi rate. Thus, our simulations
show that in 3-D it may be even more difficult than in 2-D to explain
the inactive mode of accretion. However, if the torus
forms then our simulations show that the torus will precess.
This precession may have important consequences in terms of reducing
the mass accretion rate. Namely, a precessing torus may
produce a precessing jet/wind that will then affect a larger
volume of the surrounding material than a non-precessing jet/wind.

Our simulations are in principle relevant
to any type of a black hole, which accretes gas with some 
small angular momentum, because the results should 
scale with the black hole mass. Note, that the parameters in our
models were chosen so that the ratio of the Bondi radius to the
Schwarzschild radius was equal to 10$^{3}$ and the computational domain
was between 1.5 $R_{\rm S}$ and 1.2 R$_{\rm B}$.
In this context, one of the most interesting cases is that of
the low luminosity active galaxies, and in particular the Sgr A$^{*}$,
for which the supermassive black hole of $M\sim 3.5\times 10^{6} M_{\odot}$ 
was identified in the center (Ghez et al. 2005).

The Bondi accretion rate in the Galaxy center inferred from the studies of 
stellar winds was estimated to be between $10^{-6}$ and 
$10^{-4}$ $M_{\odot}$yr$^{-1}$, while the polarization studies suggest that 
the actual accretion rate is between $10^{-7}$ and $10^{-5}$ 
$M_{\odot}$yr$^{-1}$ from ROSAT observations 
(Quataert, Narayan \& Reid 1999; Baganoff et al. 2003; Bower et al. 2005).
Although the values overlap, most of the authors agree that in Sgr A$^{*}$
the accretion rate onto the black hole is well below the Bondi value. 
Also, studies of other quiescent AGN suggest that some modification of 
the Bondi accretion is needed  (e.g. Di Matteo et al. 2000).

The hydrodynamical studies of black hole accretion in Sgr A$^{*}$ which took 
into account a random distribution of stars were presented in Coker \& Melia
(1997). The analytical estimates of the effective angular momentum
of the accreting gas performed by Mo\'scibrodzka, Das \& Czerny (2006) 
were based on the strengths of the stellar winds.
Recently, Cuadra, Nayakshin \& Martins (2006) modeled the wind accretion 
allowing the stars to move on the elliptical orbits.
We show here that 
the accretion rate of less than 40\% of the
Bondi value is possible also in models, where the
rotationally supported torus does not close, and an asymmetric
structure of material being almost spherical on one side of the black
hole may persist as a kind of steady-state (model with 
$\Delta \phi_{0} = 120^{\circ}$). Only for very small input 
of the angular momentum gas, corresponding to the total 
$\Delta \phi_{0} \le 60^{\circ}$, the accretion flow is still remains
almost spherical and the accretion rate does not drop much 
below the Bondi value.

Perhaps our most intriguing results we found, is the instability and 
precession of the torus. The precession occurs for
the closed rotationally supported torus, which forms for 
large angular momentum input $330^{\circ} \leq \Delta \phi_{0} \leq
360^{\circ}$. In other words, we found that even a very small asymmetry
in the angular momentum distribution (not necessarily in the initial
conditions) will lead to the torus
misplacement from the equatorial plane and its precession.
This happens after a few tens of orbital cycles.
For the models where the torus did not form due to a very large 
content of non rotating gas accreting in purely radial direction,
the asymmetric condition makes the quasi-steady configuration very
unstable, and rapid fluctuations of the global flow pattern occur
on the timescales of a few dynamical cycles. The clumps of gas with
large density become misplaced from the equator. For example, in 
the model with $\Delta \phi_{0} = 240^{\circ}$ 
the misplacement reaches a few tens of degrees and changes 
in the flow configuration are extremely violent.

The problem of the stability of accretion tori with respect to the
axial perturbations was studied in a number of papers.
The classical Rayleigh condition for the torus stability
(sufficient only for axisymmetric modes)  is 
that the specific angular momentum should not decrease outwards
(see e.g. Chandrasekhar 1961). The Kelvin-Helmholtz instability 
occurs when two superimposed layers of fluid are in a relative
motion. 
When the velocity shear exceeds a critical value, the resulting
pressure 
gradient (from Bernoulli's law) between the peaks and troughs of an 
interfacial wave overcomes the surface tension and gravity, and  the 
mode grows exponentially. 

Papaloizou \& Pringle (1984) studied the stability of the non-axisymmetric 
modes of the differentially rotating tori. In this first paper, 
they limited their considerations to the homentropic tori with constant 
specific angular momentum, and they found, that all such tori are unstable 
to the low order modes and the instability occurs on a dynamical timescale.
These instabilities are found to be global, i.e. their presence cannot
be detected via the local analysis nor from the considerations of
axisymmetric modes. 
In their second paper (Papaloizou \& Pringle 1985), they 
considered the tori with non-constant specific angular momentum and 
they found that for low azimuthal number $m$
the instability is driven by a Kelvin-Helmholtz mechanism.
The modes are stable in disks with angular velocity decreasing with
radius as $\Omega \propto r^{-q}$, if $q>\sqrt{3}$. 
For high $m$, the modes regain their sonic character, 
and there exist sonic modes which are driven by both mechanisms. 
However, their analytical calculations were done in the limit of a 
Keplerian disk ($q=3/2$). 
In our models, $q$ is about $1.8$. In the innermost parts, the flow is 
supersonic and compressible. Future work is needed to perform
stability analysis for such flow properties.

From the point of view of numerical, multi-dimensional simulations, as 
it has been recently discussed by
Foglizzo, Galletti \& Ruffert (2005) that it is not always obvious to determine whether
the hydrodynamical instabilities are a physical or numerical effect.
As shown by our simulations, and confirmed by the tests with smaller 
Mach numbers (models $M_{100}$ and $M_{300}$),
the instabilities that lead to the torus precession do not develop in a subsonic flow (see also Mo\'scibrodzka \& Proga 2008).
The acoustic instability develops in the 3-D accretion when the Mach number is large,
and for $\gamma=5/3$, and it is found to be a physical instability in a 
strongly supersonic flow. 
We plan to verify this result for other values of $\gamma$, and
perform more detailed resolution tests in the future work.

From the observational point of view, the precessing torus might be relevant 
to the interpretation of jet emitting sources, provided that the jet axis is
always perpendicular to the disk surfaces.
Recent observations of radio jets 
show jet reorientation.  One of possible explanation 
for this reorientation may be the 
jet precession, as was suggested e.g., for
the shape of the source 3C294 (Erlund et al. 2006). 
Also, the morphology of the BAL quasar 1045+352 indicates 
either a  precessing jet or an ongoing merger process
(Kunert-Bajraszewska \& Marecki 2007).
In addition, the morphology of some of the VLBA observed  curved jets
suggests jet precession because of
relatively short timescales (Lister 2006).

\acknowledgements
We thank  Monika Mo\'scibrodzka and Aneta Siemiginowska for helpful 
discussions and comments. We thank the developers of ZEUS-MP
for providing the code publicly available and for support.
This work was supported by 
NASA under grant NNG05GB68G.
This work was also supported by the 
 National Science Foundation through TeraGrid resources provided by
National Center for Supercomputing
Applications under grant AST070019N.

\clearpage

\begin{table}
\caption{Summary of the models for non-axisymmetric accretion
  hydrodynamics in 3-D}
\begin{center}
\begin{tabular}{l c c c c c c c r}     
\hline\hline
 Model & $\Delta \phi_{0}$ & Resolution & $T_{\rm end}$  & $\dot M$  \\ 
       & [$^{\circ}$]  & [$N_{\rm r} \times N_{\theta} \times N_{\phi}$] 
 & [$t_{\rm orb}(R_{\rm B})$] & [$\dot M_{\rm B}$]  & Presence of torus & Precession\\
\hline 
$A_{32}$ & 330 & 140x96x32 & 0.36 & 0.25 & yes & yes \\
$A_{60}$ & 330 & 140x96x60 & 0.20 & 0.25 & yes & yes \\
$B_{60}$ & 240 & 140x96x60 & 0.30 & 0.28 & not closed & --  \\
$C_{32}$ & 120 & 140x96x32 & 0.36 & 0.37 & not closed & -- \\
$C_{60}$ & 120 & 140x96x60 & 0.16 & 0.35  & not closed & -- \\
$D_{32}$ & 60  & 140x96x32 & 0.18 & 0.90  & no & -- \\
$E_{32}$ & 30  & 140x96x32 & 0.14 & 0.98 & no & -- \\
$R_{10}$  & 360 & 140x100x10 & 0.32 & 0.24 & yes & no \\
$R_{32}$ & 360 & 140x96x32  & 0.36 & 0.24 & yes & yes \\
$R_{60}$ & 360 & 140x96x60  & 0.36 & 0.24  & yes & yes \\
$M_{100}$ & 360 & 140x96x32  & 11 & --  & yes & no \\
$M_{300}$ & 360 & 140x96x32  & 2.19 & --  & yes & no  \\
\hline
\end{tabular}
\end{center}
\label{tab:models}
\end{table}

\clearpage

\voffset=-8 cm

\hoffset=0 cm
\voffset=0 cm

\clearpage

\hoffset=0 cm
\voffset=0 cm

\voffset=0 cm
\hoffset=0 cm

\clearpage

\voffset=0 cm
\hoffset=0 cm

\clearpage


\voffset=0 cm

\thispagestyle{empty}
\voffset=-2 cm


\voffset=0 cm


\end{document}